\documentclass[twocolumn,showpacs,preprintnumbers,amsmath,amssymb,pre]{revtex4-1}

\usepackage{rotating}
\usepackage{amssymb}
\usepackage{amsmath}
\usepackage{graphicx,psfrag}
\usepackage{graphics,psfrag}

\newcommand{\m}{m_{r}}
\newcommand{\tr}{\mathrm{tr}}
\def\eref#1{(\ref{#1})}

\begin{document}

\title{A non-perturbative renormalization group study of the
  stochastic Navier--Stokes equation}

\author{Carlos Mej\'ia-Monasterio}
\affiliation{Laboratory of Physical Properties, Department of Rural Engineering,
Technical University of Madrid, Av. Complutense s/n, 28040 Madrid, Spain.}
\email{carlos.mejia@upm.es}
\author{Paolo Muratore-Ginanneschi}
\affiliation{University of Helsinki, Department of Mathematics and Statistics
    P.O. Box 68 FIN-00014, Helsinki, Finland}
\email{paolo.muratore-ginanneschi@helsinki.fi}

\begin{abstract}
  We study the renormalization group flow of the average action of the
  stochastic  Navier--Stokes equation with power-law  forcing. Using
  Galilean invariance  we introduce a  non-perturbative approximation
  adapted to the zero frequency sector of the theory in the parametric
  range  of the  H\"older exponent $4-2\,\varepsilon$ of  the forcing
  where real-space local interactions are relevant.    In any
  spatial dimension $d$, we observe the convergence of the resulting  
  renormalization group flow to a unique fixed point which yields a 
  kinetic energy spectrum scaling in agreement with canonical dimension 
  analysis. Kolmogorov's $-5/3$ law is, thus, recovered for $\varepsilon=2$
  as also predicted by perturbative renormalization. At variance with
  the perturbative prediction, the $-5/3$ law emerges in the presence
  of a \emph{saturation} in the $\varepsilon$-dependence of the scaling 
  dimension  of the eddy diffusivity at $\varepsilon=3/2$ when, 
  according to perturbative renormalization, the velocity field becomes 
  infra-red relevant.
\end{abstract}

\maketitle

Kolmogorov's  K41  theory  \cite{Ko41,Ko41a}  is  the  cornerstone  of
current  understanding  of  fully  developed turbulence  in  Newtonian
fluids. A modern formulation of the theory \cite{Frisch95} is based on
the  asymptotic  solution  of the  K\'arm\'an-Howarth-Monin  equation,
expressing    energy    balance,    for   stochastic    incompressible
Navier--Stokes equation
\begin{eqnarray}
\label{intro:NS}
(\partial_{t}+\boldsymbol{v}\cdot\partial_{\boldsymbol{x}}
-\kappa\partial_{\boldsymbol{x}}^{2})\boldsymbol{v}=
\boldsymbol{f}-\partial_{\boldsymbol{x}}P \ ,
\end{eqnarray}
with   $\boldsymbol{f}$   Gaussian,   incompressible,   zero   average
time-decorrelated with correlation
\begin{eqnarray}
\label{intro:cor}
\prec\,\boldsymbol{f}(\boldsymbol{x}_{1},t_{1})\otimes \boldsymbol{f}
(\boldsymbol{x}_{2},t_{2})\,\succ
=\delta(t_{12})\,\mathsf{F}(\boldsymbol{x}_{12}) \ .
\end{eqnarray}
Here  $\prec\,\,\succ$  denotes the  ensemble  average, $\otimes$  the
tensor                                                         product,
$\boldsymbol{x}_{ij}:=\boldsymbol{x}_{i}-\boldsymbol{x}_{j}$,
$t_{ij}:=t_{i}-t_{j}$   and   $P$  is   a   pressure  term   enforcing
incompressibility:    $\partial_{\boldsymbol{x}}\cdot\boldsymbol{v}=0$.
The  solution  of K\'arm\'an-Howarth-Monin  equation  predicts in  any
spatial dimension strictly larger than two that the energy injected by
the  external  stirring ($\boldsymbol{f}$)  around  a typical  spatial
scale  $L$  is \emph{conserved}  across  an  \emph{inertial range}  of
scales  through  a  constant-flux  transfer  mechanism,  the  ``energy
cascade'',  before being  dissipated by  molecular viscosity.   In two
dimensions,  energy  and enstrophy  conservation  across the  inertial
range calls  for a  distinct analysis of  the K\'arm\'an-Howarth-Monin
equation  \cite{Be99,Be00,Li96} formalizing  the  ideas introduced  by
Kraichnan  in  \cite{Kr67}.  The  solution  predicts  a constant  flux
inverse  energy cascade  from the  injection scale  towards  the fluid
integral scale.   Below the injection scale a  constant flux enstrophy
cascade   towards  the   dissipative   scale  may   take  place   (see
e.g. \cite{Bo07}). The very  existence and properties of the enstrophy
cascade are, however, sensitive  to the boundary conditions imposed on
(\ref{intro:NS}) and  the eventual presence  and shape of  large scale
friction    mechanisms   \cite{NaAnGuOt99,Be99,CoRa07}.    Dimensional
considerations based  on the solution  of the K\'arm\'an-Howarth-Monin
equation lead  then to scaling predictions  for statistical indicators
of the flow, including the $-5/3$ exponent for the $3d$ kinetic energy
spectrum.  These predictions convincingly  account for a wide range of
experimental     and      numerical     observations     (see     e.g.
\cite{Frisch95,Falkovich}    and     references    therein).     Their
first-principle  derivation  is  therefore  a  well-grounded  research
question.
A useful  tool to pursue this  goal is offered  by the renormalization
group,  although  its application  to  the  inquiry of  Navier--Stokes
turbulence  is ridden by  challenges.  Renormalization  group analysis
\cite{FoNeSt76,FoNeSt77,DeMa79}  can  be  applied  only far  from  the
turbulent regime and for a  very special choice of the random Gaussian
field  $\boldsymbol{f}$.  This  latter needs  to have  in  any spatial
dimension   $d$   a   power-law   spectrum  with   H\"older   exponent
$4-2\,\varepsilon$:
\begin{subequations}
\label{intro:forcing}
\begin{eqnarray}
\lefteqn{
\mathsf{F}(\boldsymbol{x}_{12};m,M)
=
}
\nonumber\\&&
\int_{\mathbb{R}^{d}}\frac{d^{d}p}{(2\,\pi)^{d}}\frac{e^{\imath\,\boldsymbol{p}\cdot\boldsymbol{x}_{12}}\,
\,\mathsf{T}(\boldsymbol{p})}{d-1}
\,\check{F}(\boldsymbol{p};m,M) \ ,
\end{eqnarray}
\begin{eqnarray}
\check{F}(\lambda\,\boldsymbol{p};\lambda\, m\,,\lambda\, M)=
\lambda^{4-d-2\varepsilon}\check{F}(\boldsymbol{p};m,M) \ ,
\end{eqnarray}
\end{subequations}
with       $\mathsf{T}(\boldsymbol{p})=\mathsf{I}-\boldsymbol{p}\otimes
\boldsymbol{p}/p^{2}$         the        transverse        projector,
$p:=\parallel\boldsymbol{p}\parallel$, and $m  \ll M$ respectively the
inverse  integral  and  ultra-violet   scales  of  the  forcing.   The
rationale  for the  choice  is that  for  vanishing $\varepsilon$  the
canonical      scaling      dimensions      of     the      convective
(i.e.                 $\partial_{t}\boldsymbol{v}$,                and
$\boldsymbol{v}\cdot\partial_{\boldsymbol{x}}\boldsymbol{v}$)       and
dissipative (i.e. $\partial_{\boldsymbol{x}}^{2}\boldsymbol{v}$) terms
in  the Navier--Stokes  equation tend  to  the same  value. This  fact
suggests   that  for  $\varepsilon$   equal  zero   canonical  scaling
dimensions  may coincide with  the exact  scaling dimensions.  In this
sense,  the  vanishing  $\varepsilon$  case defines  a  \emph{marginal
  scaling   limit}   around  which it may be possible to determine   
scaling dimensions by means of a perturbative expansion in $\varepsilon$ 
in analogy  to what is done for  critical phenomena  described by 
a Boltzmann  equilibrium (see e.g. \cite{Zinn,Cardy96}).  
For the stochastic Navier--Stokes equation the  situation  is,  however,  
not conclusive.  Renormalization  group yields in any spatial dimension a 
kinetic energy spectrum
\begin{eqnarray}
\label{intro:rgspectrum}
\mathcal{E}(p)\propto p^{\eta_{2:0}}\,,\hspace{1.0cm}\eta_{2:0}=1-4\varepsilon/3
\end{eqnarray}
\cite{DeMa79} (see also \cite{AdAnVa99} for an exhaustive review). In
(\ref{intro:rgspectrum}) the exponent labeling emphasizes the
possibility of sub-leading corrections.  Fully developed turbulence in
$3d$ should correspond to an infra-red dominated spectrum of the
stirring force as it occurs for $\varepsilon\geq 2$.  Interestingly,
(\ref{intro:rgspectrum}) recovers Kolmogorov's result for
$\varepsilon$ equal two.  Consistence with Kolmogorov theory then
requires the exponent in (\ref{intro:rgspectrum}) to freeze for
$\varepsilon$ larger than two to the value $-5/3$.  Within the
perturbative renormalization group framework, the occurrence of such
non-analytic behavior can only be argued \cite{FoFr83}.  Direct
numerical simulations \cite{SaMaPa98,BiCeLaSbTo04} exhibited, within a
$512^{3}$-lattice accuracy, a transition in the
$\varepsilon$-dependence of $\eta_{2}$ which is consistent with the
freezing scenario.  The situation is, however, completely different in
two dimensions \cite{MaMGMu07,MaMGMu09}.  On the one hand,
perturbative renormalization group analysis \cite{Ho98} upholds the
validity of (\ref{intro:rgspectrum}) for any $\varepsilon$.  On the
other hand, the asymptotic solution of the K\'arm\'an-Howarth-Monin
equation \cite{MaMGMu09} shows that (\ref{intro:rgspectrum}) is always
sub-dominant with respect to the inverse energy cascade spectrum
$\mathcal{E}(p)\propto p^{-5/3}$, for $\varepsilon\leq 2$ i.e.
\emph{even} in the regime where renormalization group analysis should
apply.  Direct numerical simulations up to $2048^{2}$ resolution give
clear evidence of the inverse cascade \cite{MaMGMu07,MaMGMu09}.  A
scenario reconciling these findings may be that the
Kraichnan-Kolmogorov inverse cascade corresponds to a renormalization
group non-perturbative fixed point which does not bifurcate from the
Gaussian fixed point at marginality.  Evidences of the occurrence of
such an ``exotic'' phenomenon, have been given in models of wetting
transitions by ``non-perturbative approximations'' of the Wilsonian
renormalization group \cite{LiFi86,LiFi87}.  More recently, similar
methods gave evidence of the existence of a strong coupling fixed
point in the Kardar-Parisi-Zhang model of interfacial growth
\cite{CaChDeWs09}, yielding scaling predictions favorably comparing
with direct numerical simulations.  Motivated by these results, in the
present contribution we derive the exact renormalization group
equations for the stochastic Navier--Stokes equation. We then
investigate them using a ``non-perturbative approximations'' similar
to the one used in \cite{CaChDeWs09}.  By this we mean, as often done
in non-perturbative renormalization \cite{BeTeWe02,BaBe01},
truncations of the flow equations based on some assumption on the
physical properties of the inquired system.  Specifically, we
investigate the consequences of the simplest closure compatible with
Galilean invariance and with the number of relevant interactions
identified by perturbative renormalization at small $\varepsilon$.
The second requirement guarantees the existence of a limit where the
closure becomes exact in the sense that it recovers the perturbative
renormalization group fixed point.  As in \cite{CaCh07,CaChDeWs09}, we
focus on the exact renormalization group equations for the
\emph{average action} or \emph{thermodynamic potential} defined by the
stochastic Navier--Stokes equations. In striking contrast with the
compressible stochastic dynamics studied in \cite{CaChDeWs09}, we do
not find any evidence of a non-perturbative fixed point which may be
associated to constant flux solutions in general and to the two
dimensional inverse cascade in particular.  The truncation we consider
reproduces instead the expected correct scaling behavior in the regime
dominated by real-space local interactions i.e. $d=3$ and
$\varepsilon\,\leq 3/2$.  Interestingly, we observe in any dimension a
transition at $\varepsilon=3/2$ in the scaling behavior of the
eddy-diffusivity. This latter deviates from the renormalization group
scaling prediction by freezing from there on in $\varepsilon$ to its
$\varepsilon=3/2$ value.  This result was previously derived by
different methods in \cite{MoWe95}. It is worth noticing that  $\varepsilon=3/2$ 
is the threshold value after which the critical dimension of the 
stochastic Navier--Stokes velocity predicted by perturbative renormalization
becomes negative or, in other words, infra-red relevant.
In spite of the eddy diffusivity
saturation, we obtain a kinetic energy spectrum scaling in agreement
to (\ref{intro:rgspectrum}) with no saturation for
$\varepsilon\,>\,2$. This latter fact is not entirely surprising since
the particle irreducible vertices contributing to the approximated
renormalization group flow, are only a subset of those needed to fully
reconstruct the flux i.e. the chief statistical indicator in
Kolmogorov's theory.  

The structure of the paper is as follows.  In
section~\ref{sec:KHM} we briefly recall the K\'arm\'an-Howarth-Monin
equation and its predictions for power law forcing. In
section~\ref{sec:RGEQ} we derive the exact renormalization group
average action 
for the model. The scope of these sections is to provide basic
background on turbulence and functional renormalization to facilitate
the reading by researchers familiar with one of these subjects but not
the other.  Using the Ward identities imposed by Galilean and
translational invariance in section~\ref{sec:approx} we introduce our
approximations of the exact flow.  We write the resulting equations in
section~\ref{sec:eqs} where we also outline their qualitative
analysis.  To simplify the discussion we detail auxiliary formulas in
appendix~\ref{ap:cvs}.  An advantage of our formalism is that by
preserving the structure of the exact renormalization group flow it
guarantees the ``realizability'' of the ``closure'' that we impose
\cite{BoKrOt93}.  In \ref{sec:toy} we describe the analytic solution
of our equations in a simplified limit.  Section~\ref{sec:numerics}
reports the result of the numerical integration of our equations
respectively in the three and two-dimensional cases.  Finally we turn
in~\ref{sec:concl} to discussion and conclusions.

\section{Scaling predictions based on the K\'arm\'an-Howarth-Monin equation}
\label{sec:KHM}

The K\'arm\'an-Howarth-Monin equation describes the energy balance in
the putative unique steady-state to which Galilean invariant
statistical indicators are expected to converge. Specifically, if we
consider the two-point equal time correlation tensor
\begin{eqnarray}
\label{KHM:2p}
\mathsf{C}_{2}(\boldsymbol{x}_{12},t)=
\prec\,\boldsymbol{v}(\boldsymbol{x}_{1},t)\otimes\boldsymbol{v}(\boldsymbol{x}_{2},t)\,\succ \ ,
\end{eqnarray}
and the three point equal time structure tensor
\begin{eqnarray}
\label{KHM:3p}
&&\hspace{-0.5cm}\mathsf{S}_{3}(\boldsymbol{x}_{12},t)=
\prec\,\delta \boldsymbol{v}\left(\boldsymbol{x}_{12},t\right)\otimes
\delta \boldsymbol{v}\left(\boldsymbol{x}_{12}\right)\otimes\delta \boldsymbol{v}\left(\boldsymbol{x}_{12}\right)
\succ \ ,
\\
\label{KHM:increment}
&&\hspace{-0.5cm}
\delta \boldsymbol{v}\left(\boldsymbol{x}_{12}\right)
:=\boldsymbol{v}(\boldsymbol{x}_{1},t)-\boldsymbol{v}(\boldsymbol{x}_{2},t) \ ,
\end{eqnarray}
a straightforward calculation using incompressibility and the inertial range 
translational and parity invariance yields 
\begin{eqnarray}
\label{KHM:KHM}
\partial_{t}C+\frac{1}{2}\partial_{\boldsymbol{x}}\cdot\boldsymbol{S}
-2\,\kappa\,\partial_{\boldsymbol{x}}^{2}C=F \ ,
\end{eqnarray}
for $C:=\tr \,\mathsf{C}_{2}$, $F:=\tr \,\mathsf{F}$ and
$S^{\alpha}:=\mathsf{S}_{3\hspace{0.3cm}\alpha_{1}}^{\alpha
  \alpha_{1}}$ and Einstein convention on repeated indices. In any
spatial dimension strictly larger than two, (\ref{KHM:KHM}) admits an asymptotic
solution under the hypotheses (see e.g. \cite{Frisch95} for a detailed
discussion) that \emph{(i-1)} statistical indicators attain a unique
steady state and hence $\partial_{t}C=0$, \emph{(ii-1)} they are
smooth for any finite molecular viscosity but \emph{(iii-1)} the
inviscid limit of the energy dissipation exhibits a dissipative
anomaly
\begin{eqnarray}
\label{KHM:da}
0\,<\,-2\,\lim_{\kappa\downarrow 0}\lim_{\parallel\boldsymbol{x}\parallel\downarrow 0}\kappa\,\partial_{\boldsymbol{x}}^{2}C\neq
-2\,\lim_{\parallel\boldsymbol{x}\parallel\downarrow 0}\lim_{\kappa\downarrow 0}\kappa\,\partial_{\boldsymbol{x}}^{2}C=0 \ .
\end{eqnarray}
Under these hypotheses, if the dominant contribution to the forcing 
correlation comes from wave numbers of the order $m$,  
Kolmogorov's classical result \cite{Ko41}
\begin{eqnarray}
\label{KHM:KL}
\lefteqn{
\lim_{\parallel\boldsymbol{x}\parallel\downarrow 0}\lim_{\kappa\downarrow 0}
\partial_{x^{\beta}}\mathsf{S}_{3}^{\alpha_{1}\alpha_{2}\alpha_{3}}(\boldsymbol{x})
} \ ,
\nonumber\\&&
=-\,\frac{2\, \bar{E}}{d\,(d+2)} \mathcal{P}_{\boldsymbol{\alpha}}
\left\{ \delta^{\beta\alpha_{1}}\delta^{\alpha_{2}\alpha_{3}}\right\} \ ,
\end{eqnarray}
holds true for $m\parallel\boldsymbol{x}\parallel\,\ll\,1$,
$\mathcal{P}_{\boldsymbol{\alpha}}$ being the index cyclical
permutation operation over
$\boldsymbol{\alpha}=(\alpha_{1},\alpha_{2},\alpha_{3})$ and, in accordance with Kolmogorov's notation \cite{Ko41,Ko41a}, 
$\bar{E}=F(0)/2$ the mean dissipation of energy . In other
words, the leading scaling exponent of (\ref{KHM:3p}) is
\begin{eqnarray}
\label{}
\zeta_{3:0}=1 \ .
\end{eqnarray} 
Dimensional considerations then yield for the kinetic energy spectrum
scaling exponent the Kolmogorov's scaling law
\begin{eqnarray}
\label{KHM:Kspectrum}
\eta_{2:0}=-5/3 \ .
\end{eqnarray}
If instead the forcing correlation is a power-law within the range of
scales $M^{-1}\ll\parallel\boldsymbol{x}\parallel\ll m^{-1}$ with
H\"older exponent $4-2\,\varepsilon$, we should distinguish two
situations.  If $\varepsilon\,<\,2$, the forcing correlation
(\ref{intro:forcing}) remains well-defined in the limit of infinite
integral scale $m^{-1}$. In such a case (\ref{KHM:KL}) holds for
$M^{-1}\ll\parallel\boldsymbol{x}\parallel\,\ll\,\ell$ where we
introduced $\ell=\kappa\,/\sqrt{ F(0)}$, the typical scale below which
molecular dissipation dominates. Under the present hypotheses
$\ell\propto \kappa\,M^{\varepsilon-2}$, the omitted proportionality
factor being a dimensional constant independent of $\kappa$ and
$M$. This range of scales is not accessible by perturbative
ultra-violet renormalization group methods. These latter may describe
instead the range $\parallel\boldsymbol{x}\parallel\,\gg\,M^{-1}$
where the asymptotic solution of (\ref{KHM:KHM}) states that the
leading scaling exponent of (\ref{KHM:3p}) is
\begin{eqnarray}
\label{KHM:3dls}
\zeta_{3:0}=-3+2\varepsilon \ .
\end{eqnarray}
Dimensional analysis based on (\ref{KHM:3dls}) then recovers the
renormalization group prediction (\ref{intro:rgspectrum}) for the
kinetic energy spectrum. A different scenario occurs for
$\varepsilon\,>\,2$: the forcing correlation has a finite limit if the
ultra-violet scale $M$ tends to infinity for any finite value of the
inverse integral scale $m$. In the range
$m^{-1}\,\ll\,\parallel\boldsymbol{x}\parallel\,\ll\,\ell$,
$\ell\propto \kappa\,m^{\varepsilon-2}$, (\ref{KHM:KL}) holds with
possible sub-dominant terms with scaling dimension (\ref{KHM:3dls}).
To summarize, the hint coming from the K\'arm\'an-Howarth-Monin
equation for spatial dimensions $d\,>\,2$ is that the $-5/3$ exponent
stems from the dominance for $\varepsilon\,>\,2$ of the constant flux
over the dimensional scaling asymptotic solution of
(\ref{KHM:KHM}). This result can be justified within perturbative
renormalization theory using an argument proposed by Fournier and
Frisch in \cite{FoFr83} (see also \cite{AdAnVa99}). It is worth here
to briefly recall this argument in order to evince the assumptions on
which it relies.  Let $\check{C}(\boldsymbol{p})$ be the Fourier transform
of the trace of the two-point equal time correlation tensor
(\ref{KHM:2p}). Renormalization group analysis upholds that the
expansion in powers of $\varepsilon$ can be re-summed in the form
\begin{eqnarray}
\label{KHM:rgresum}
\check{C}(p)\overset{M\uparrow\infty}{\to}\nu^{2}(p)\,p^{2-d}\,c\left(\frac{m}{p},\varepsilon\right)
\end{eqnarray}
Here, $c$ is a function independent of $M$ which can be determined order by order
in a regular expansion around the renormalized theory 
(the so called ``renormalized perturbation theory''). The $\nu(p)$ in the prefactor
is the ``\emph{running viscosity}'' the explicit form whereof within all orders in $\varepsilon$ 
is the main achievement of renormalization group analysis: 
\begin{eqnarray}
\label{KHM:runvis}
\nu(p)=\left[\frac{F(0)\,m_{\star}^{2\,\varepsilon-4}}{p^{2\,\varepsilon}}\right]^{1/3}
\end{eqnarray}
A crucial role here is played by the mass scale $m_{\star}$. Since for 
$\varepsilon\,<\,2$ the theory is well-defined in the limit of infinite integral 
scale, $m_{\star}$ in this range must have a finite limit as $m$, the inverse integral scale, 
tends to zero. For $\varepsilon\,>\,2$, on the contrary, the energy input becomes infra-red dominated 
and as a consequence $m_{\star}\propto m$. Finally, let us observe 
following \cite{FoFr83,AdAnVa99} that comparison
with Kolmogorov theory should be done by \emph{holding fixed} the energy input while
taking the limit of infinite integral scale. Let us \emph{assume} that: \textbf{A} the resummation 
(\ref{KHM:rgresum}) holds for any finite $\varepsilon$ and, \textbf{B} (\ref{KHM:rgresum})
admits a finite limit as the integral scale $m^{-1}$ tends to infinity. 
Under these hypotheses it follows immediately that
\begin{eqnarray}
\label{KHM:rghyp}
\lim_{\substack{m\downarrow 0\\ F(0)=\mathrm{constant}}}\lim_{M\uparrow\infty}C(p)\sim
\left\{
\begin{array}[c]{ll}
p^{2-4\varepsilon/3-d} \hspace{0.2cm}&\hspace{0.2cm}\varepsilon\,<\,2
\\
p^{2-8/3-d} \hspace{0.2cm}&\hspace{0.2cm}\varepsilon\,>\,2
\end{array}
\right.
\end{eqnarray}
Note that (\ref{KHM:rghyp}) is equivalent to say that $c$ is finite in the limits
for $\varepsilon\,<\,2$ and divergent for $\varepsilon\,>\,2$.
Two mechanisms may obviously invalidate this result.
Assumption \textbf{A} breaks down if for some finite $\varepsilon$ 
a new fixed point of the renormalization group transformation appears.
This may lead to a different result for the \emph{running viscosity} 
(\ref{KHM:runvis}) marking the onset of a different critical regime.
Glazek and Wilson gave in \cite{GlWi04} an analytically tractable example of a 
non-perturbative bifurcation of renormalization group flow fixed point.
Scenarios for the break-down of \textbf{A} were already discussed in \cite{DeMa79}.
Checking the validity of assumption \textbf{B} requires controlling the function 
$c$ in (\ref{KHM:rgresum}) in the limit of vanishing $m$. The needed technical tool is 
the so-called operator-product-expansion \cite{Zinn,Cardy96}. 
In particular, $c$ may become divergent for $m\downarrow 0$ 
above some threshold value $\varepsilon_{\star}\,<\,2$ as some irrelevant 
composite operator contributing to $\check{C}$ turns relevant. Examples of such operators are known
\cite{AdAnVa99}: the velocity field and its integer powers become
relevant at $\varepsilon=3/2$ the energy dissipation at $\varepsilon=2$.
In summary, the domain of validity in $3d$ of the renormalization group
predictions and, even more, the exponent ``freezing'' needed to
recover Kolmogorov theory are open research questions which we set out
explore in the present contribution.

The asymptotic analysis of (\ref{KHM:KHM}) in $2d$ must be treated
apart in order to take into account enstrophy conservation. In
particular \cite{Be99,Be00}, Kraichnan's theory \cite{Kr67} is
epitomized by a more restrictive version of \emph{(i-1)}, which we will
refer to as \emph{(i-2)}, requiring only Galilean invariant quantities
to reach a steady state. In other words, $\partial_{t}C$ does not
vanish.  Furthermore, \emph{(iii-1)} is replaced by a new hypothesis
\emph{(iii-2)} ruling out the occurrence of dissipative anomaly for the
kinetic energy dissipation:
\begin{eqnarray}
\label{KHM:noda}
\lim_{\kappa\downarrow 0}\lim_{\parallel\boldsymbol{x}\parallel\downarrow 0}\kappa\,\partial_{\boldsymbol{x}}^{2}C
=\lim_{\parallel\boldsymbol{x}\parallel\downarrow 0}\lim_{\kappa\downarrow 0}\kappa\,\partial_{\boldsymbol{x}}^{2}C=0 \ .
\end{eqnarray}
It is worth noticing that \emph{(iii-2)} can be rigorously proved to
hold true in some setup for the deterministic Navier--Stokes
\cite{CoRa07} (see also discussion in \cite{Ku09}).  We refer the
reader to \cite{MaMGMu09} for a detailed analysis of the
two-dimensional K\'arm\'an-Howarth-Monin equation in the power-law
case also corroborated by direct numerical simulations of
(\ref{intro:NS}). Here we only summarize the results.  In the range of
scales which can be investigated  by perturbative ultra-violet
renormalization group methods, three distinct regimes may
set in depending upon the value of $\varepsilon$.  For
$\varepsilon\,<\,2$, the ultra-violet cut-off gives the dominant contribution to 
the total energy $F(0)\propto
M^{4-2\,\varepsilon}$ and enstrophy $-(\partial_{\boldsymbol{x}}^{2}F)(0)\propto M^{6-2\,\varepsilon}$. 
Correspondingly, the inviscid limit in the range
$M\,\parallel\boldsymbol{x}\parallel\,\gg\,1$ predicts for the leading
and sub-leading scaling exponents of (\ref{KHM:3p})
\begin{eqnarray}
\label{KHM:inverse}
\zeta_{3:0}=1\hspace{1.0cm}\&\hspace{1.0cm}\zeta_{3:1}=-3+2\,\varepsilon \ .
\end{eqnarray}
This is in agreement with Kraichnan's theory which predicts the onset
of an inverse energy cascade for wave-numbers smaller than the one
characteristic of the (total) input.  The ensuing dimensional
prediction for the kinetic energy spectrum scaling exponent is (\ref{KHM:Kspectrum})
while $\eta_{2:1}=1-4/3\varepsilon$ only describes a sub-leading
correction.  For $2\,<\,\varepsilon\,<\,3$, $F(0)\propto
m^{4-2\,\varepsilon}$ and $-(\partial_{\boldsymbol{x}}^{2}F)(0)\propto
M^{6-2\,\varepsilon}$ indicate that in the region
$m^{-1}\,\gg\,\parallel\boldsymbol{x}\parallel\,\gg\,M^{-1}$ the third
order structure tensor is sustained by an input of enstrophy from larger
wave-numbers and an input of energy from smaller wave-numbers. As a
result, the flux balances locally in real space with the forcing so that
(\ref{KHM:3dls}) holds true.  Finally for $\varepsilon\,>\,3$ and in
the presence of a large-scale hypo-friction \cite{MaMGMu09} both
energy $F(0)\propto m^{4-2\,\varepsilon}$ and
enstrophy$-(\partial_{\boldsymbol{x}}^{2}F)(0)\propto
m^{6-2\,\varepsilon}$ inputs are dominated by the infra-red mass
scale $m$. As a consequence, a direct enstrophy cascade sets in for
$m \parallel\boldsymbol{x}\parallel\,\ll\,1$ and
\begin{eqnarray}
\label{KHM:direct}
\zeta_{3:0}= 3\hspace{1.0cm}\&\hspace{1.0cm}\zeta_{3:1}=-3+2\,\varepsilon \ .
\end{eqnarray}
Again, dimensional analysis based on (\ref{KHM:direct}) predicts
\begin{eqnarray}
\label{}
\eta_{2:0}=-3\hspace{1.0cm}\&\hspace{1.0cm}\eta_{2:1}=1-4/3\,\varepsilon \ ,
\end{eqnarray}
with the leading scaling exponent ``freezing'' at the threshold value
attained at $\varepsilon=3$.  With these results in mind, we turn now
to the formulation of a non-perturbative renormalization group theory
with the aim of collating scaling predictions for the energy spectrum.
 
\section{Renormalization group flow for the average action}
\label{sec:RGEQ}

\subsection{Thermodynamic formalism}

For finite infra-red $m$ and ultra-violet $M$ cut-offs of the Gaussian
forcing (\ref{intro:cor}) it is reasonable to assume that the generating function
\begin{eqnarray}
\label{RGEQ:gf}
\mathcal{Z}_{(\boldsymbol{\jmath},\bar{\boldsymbol{\jmath}})}:=
\prec\,e^{\boldsymbol{\jmath}\star\boldsymbol{v}(;\boldsymbol{f}+\bar{\boldsymbol{\jmath}})}\,\succ \ ,
\end{eqnarray}
is well defined. The average in (\ref{RGEQ:gf}) is over the Gaussian
statistics of the forcing,
$\boldsymbol{v}(;\boldsymbol{f}+\bar{\boldsymbol{\jmath}})$ is the
solution of (\ref{intro:NS}) for any fixed realization of
$\boldsymbol{f}$ shifted by an arbitrary source field
$\bar{\boldsymbol{\jmath}}$, and $\star$ denotes the
$\mathbb{L}^{2}\left(\mathbb{R}^{d}\times\mathbb{R}\right)$ scalar
product
\begin{eqnarray}
\label{}
\boldsymbol{\jmath}\star\boldsymbol{v}(;\bar{\boldsymbol{\jmath}})
:=\int_{\mathbb{R}^{d}\times\mathbb{R}}\hspace{-0.4cm}d^{d}x\,dt\,
\boldsymbol{\jmath}\left(\boldsymbol{x},t\right)\cdot
\boldsymbol{v}\left(\boldsymbol{x},t;\bar{\boldsymbol{\jmath}}\right) \ .
\end{eqnarray} 
Functional derivatives at zero external sources
$\left(\boldsymbol{\jmath},\bar{\boldsymbol{\jmath}}\right)$ of
(\ref{RGEQ:gf}) yield the expressions of the correlation and response (to
variations of $\boldsymbol{v}$ with respect to $\boldsymbol{f}$)
tensors of any order.
The generating function of connected correlations
\begin{eqnarray}
\label{RGEQ:fe}
\mathcal{W}_{(\boldsymbol{\jmath},\bar{\boldsymbol{\jmath}})}:=
\ln\mathcal{Z}_{(\boldsymbol{\jmath},\bar{\boldsymbol{\jmath}})} \ ,
\end{eqnarray}
is equal to minus the free energy of the field theory. 
In particular, with these conventions we have
\begin{eqnarray}
\label{RGEQ:cor}
\lefteqn{
\mathsf{C}^{\alpha_{1}\alpha_{2}}(\boldsymbol{x}_{12},t_{12})\equiv\,[\mathsf{W}^{(2,0)}]^{\alpha_{1}\alpha_{2}}
\left(\boldsymbol{x}_{12},t_{12}\right)}
\nonumber\\&&
\hspace{1.5cm}:=\left.\frac{\delta^{2}\mathcal{W}_{\left(\boldsymbol{\jmath},\bar{\boldsymbol{\jmath}}\right)}}
{\delta \jmath_{\alpha_{1}}(\boldsymbol{x}_{1},t_{1})\delta\jmath_{\alpha_{2}}(\boldsymbol{x}_{1},t_{2})}
\right|_{\boldsymbol{\jmath}=\bar{\boldsymbol{\jmath}}=0} \ . \hspace{1.0cm}
\end{eqnarray}
Analogously, the second order response function is
\begin{eqnarray}
\label{RGEQ:res}
\lefteqn{
\prec\,\frac{\delta v^{\alpha_{1}}(\boldsymbol{x}_{1},t_{1})}{\delta f^{\alpha_{2}}(\boldsymbol{x}_{2},t_{2})}\,\succ
\equiv [\mathsf{W}^{(1,1)}]^{\alpha_{1}}_{\hspace{0.2cm}\alpha_{2}}\left(\boldsymbol{x}_{12},t_{12}\right)}
\nonumber\\&&
\hspace{1.5cm}:=\left.\frac{\delta^{2}\mathcal{W}_{\left(\boldsymbol{\jmath},\bar{\boldsymbol{\jmath}}\right)}}
{\delta \jmath_{\alpha_{1}}(\boldsymbol{x}_{1},t_{1})\delta\bar{\jmath}^{\alpha_{2}}(\boldsymbol{x}_{2},t_{2})}
\right|_{\boldsymbol{\jmath}=\bar{\boldsymbol{\jmath}}=0} \ . \hspace{1.0cm}
\end{eqnarray}
The Legendre transform of the free energy (\ref{RGEQ:fe}) specifies the 
\emph{average action} or the thermodynamic potential of the statistical 
field theory: 
\begin{eqnarray}
\label{RGEQ:tp}
\mathcal{U}_{(\boldsymbol{u},\bar{\boldsymbol{u}})}:=\sup_{(\boldsymbol{\jmath}\,,\bar{\boldsymbol{\jmath}})}
\left\{\boldsymbol{\jmath}\star\boldsymbol{u}+\bar{\boldsymbol{\jmath}}\star \bar{\boldsymbol{u}}
-\mathcal{W}_{(\boldsymbol{\jmath}\,,\bar{\boldsymbol{\jmath}})}\right\} \ .
\end{eqnarray}
The Legendre anti-transform of (\ref{RGEQ:tp}) reconstruct the convex
envelope of the free energy (\ref{RGEQ:fe}). In this sense the average
action may be interpreted as an ultra-violet regularization of the
theory. The average action is a functional of the fields
$\left(\boldsymbol{u},\bar{\boldsymbol{u}}\right)$, which are Legendre conjugate to the external sources
$\left(\boldsymbol{\jmath},\bar{\boldsymbol{\jmath}}\right)$ and which
as customary will be referred to as ``classical fields''.
As extensively discussed in \cite{WE91,BeTeWe02} the average action provides a convenient
starting point for non-perturbative renormalization. Dealing with it 
is conceptually equivalent to working with
the Wilsonian effective action  as done by Polchinski in \cite{Po83}. 
Namely, the  corresponding equations can in principle  be converted in
one  another by  a Legendre  transform  if one  identifies the  running
cut-off. The average action offers, as we will see below, some technical 
advantages \cite{BaBe01} which significantly simplify the formalism.

\subsection{Flow equations}

A stationary phase approximation to (\ref{RGEQ:gf}) in the weak stirring
limit $\mathsf{F}\downarrow 0$ (see appendix~\ref{ap:JDD}) yields with
logarithmic accuracy
\begin{subequations}
\label{RGEQ:tpini}
\begin{eqnarray}
\lefteqn{
\mathcal{U}_{M}\sim
}
\nonumber\\&&
\bar{\boldsymbol{u}}\star[(\partial_{t}-\kappa\, \partial_{x}^{2})\boldsymbol{u}+
\mathsf{T}(\boldsymbol{u}\cdot\partial_{\boldsymbol{x}}\boldsymbol{u})]
-\frac{\bar{\boldsymbol{u}}\star\mathsf{F}\star\bar{\boldsymbol{u}}}{2} \ ,
\end{eqnarray}
\begin{eqnarray}
\partial_{\boldsymbol{x}}\cdot\boldsymbol{u}
=\partial_{\boldsymbol{x}}\cdot\bar{\boldsymbol{u}}=0 \ .
\end{eqnarray}
\end{subequations}
The limit $\mathsf{F}=0$ describes the trivial steady state of the decaying
Navier-Stokes equation. We posit that (\ref{RGEQ:tpini}) provides the
initial condition for the renormalization group flow of the running
average action $\mathcal{U}_{\m}$. This flow describes the building up
of the exact average action $\mathcal{U}$ of (\ref{RGEQ:gf}) as a
function of an infra-red cut-off suppressing any interaction above an
infra-red scale $\m$ and recovering $\mathcal{U}$ in the limit of
vanishing $\m$. These conditions can be matched \cite{We93,BeTeWe02}
if we replace in (\ref{intro:NS}) the molecular viscosity with an
``hyper-viscous'' term, local in wave number space,
\begin{eqnarray}
\label{RGEQ:hv}
\kappa\mapsto\tilde{\kappa}:=\kappa+\kappa_{\m}\check{R}\left(\frac{p}{\m}\right) \ ,
\end{eqnarray}
with $\check{R}$ a function rapidly decaying for large values of its
argument and diverging at the origin. A convenient choice
\cite{CaChDeWs09} is
\begin{eqnarray}
\label{RGEQ:cutoff}
\check{R}(p)=\frac{1}{e^{p^{2}}-1} \ .
\end{eqnarray} 
In (\ref{RGEQ:hv}) we also introduced the ``running'' viscosity
$\kappa_{\m}$.  We will use this extra degree of freedom to
constrain the flow to satisfy a renormalization condition on the eddy
diffusivity.  As for the viscosity, we then apply an high-pass filter to
the Gaussian forcing
\begin{eqnarray}
\label{RGEQ:forcing}
\boldsymbol{f}\mapsto\boldsymbol{\tilde{f}} \ ,
\end{eqnarray}
such that
\begin{eqnarray}
\label{RGEQ:highpass}
\prec\,\boldsymbol{\tilde{f}}_{1}\otimes \boldsymbol{\tilde{f}}_{2}\,\succ
=\delta(t_{12})\,\sum_{i=0}^{1}\mathsf{F}_{(i)}\left(\boldsymbol{x}_{12};\m\right) \ ,
\end{eqnarray}
where we defined
\begin{subequations}
\begin{eqnarray}
\label{}
\mathsf{F}_{(0)}\left(\boldsymbol{x}_{12};\m\right)=\mathsf{F}\left(\boldsymbol{x}_{12};\m,\infty\right) \ ,
\end{eqnarray}
\begin{eqnarray}
\label{NLforce}
\tr \,\check{\mathsf{F}}_{(0)}\left(\boldsymbol{p};\m\right)=
F_{o}\,\m^{4-d-2\,\varepsilon}\,(d-1)\,\chi_{(0)}\left(\frac{p}{\m}\right) \ ,
\end{eqnarray}
\end{subequations}
and
\begin{subequations}
\begin{eqnarray}
\label{RGEQ:fct}
\check{\mathsf{F}}_{(1)}\left(\boldsymbol{p};\m\right)=F_{\m}\chi_{(1)}(p;\m)\,\mathsf{T}(\boldsymbol{p}) \ ,
\end{eqnarray}
\begin{eqnarray}
\label{}
\chi_{(1)}(p;\m)\,:=\,p^{2}\,e^{-\frac{p^{2}}{\m}} \ .
\end{eqnarray}
\end{subequations}
This latter term describes a local (in the infra-red or for $\m=O(M)$)
perturbation of the measure progressively suppressed as $\m$
decreases. Locality entitles us to interpret this term as a
renormalization counter-term in the sense of
\cite{HoNa96,Ho98,AdHoKoVa05}.  Again, we will use the extra freedom
introduced by $F_{\m}$ to impose a renormalization condition on the
flow.  The replacements (\ref{RGEQ:hv}), (\ref{RGEQ:forcing}) turn
(\ref{RGEQ:gf}) into a family of generating functions differentiable
with respect to the parameter $\m$.  A straightforward calculation
(see appendix~\ref{ap:gf:rg}) yields
\begin{eqnarray}
\label{RGEQ:gfflow}
\lefteqn{
\m\partial_{\m}\mathcal{Z}_{\left(\boldsymbol{\jmath},\bar{\boldsymbol{\jmath}}\right)}=
\int_{\mathbb{R}^{d}\times \mathbb{R}^{d}\times\mathbb{R}} \hspace{-1.0cm} d^{d}x_{1}d^{d}x_{2}dt\, } 
\nonumber\\&&
\hspace{-0.3cm}
\times
\left\{\frac{(\m\partial_{\m}\tilde{\mathsf{F}})^{\alpha_{1}\alpha_{2}}(\boldsymbol{x}_{12})}{2}
\frac{\delta^{2} \mathcal{Z}_{\left(\boldsymbol{\jmath},\bar{\boldsymbol{\jmath}}\right)}}
{\delta \bar{\jmath}^{\alpha_{1}}(\boldsymbol{x}_{1},t)\delta \bar{\jmath}^{\alpha_{2}}(\boldsymbol{x}_{2},t)}\right.
\nonumber\\&&
\hspace{-0.3cm}\left.
+(\m\partial_{\m}\kappa_{\m}\,R)(\boldsymbol{x}_{12})\partial_{\boldsymbol{x}_{2}}^{2}
\frac{\delta^{2} \mathcal{Z}_{\left(\boldsymbol{\jmath},\bar{\boldsymbol{\jmath}}\right)}}
{\delta \bar{\jmath}^{\alpha_{1}}(\boldsymbol{x}_{1},t)\delta \jmath_{\alpha_{1}}(\boldsymbol{x}_{2},t)}
\right\} \ .
\end{eqnarray}
Upon defining
\begin{eqnarray}
\label{RGEQ:Rm}
\lefteqn{
\boldsymbol{\mathcal{R}}(\boldsymbol{x}_{12},t_{12})
:=
}
\nonumber\\ &&
\delta(t_{12})
\begin{bmatrix}
0 & \kappa_{\m}\mathsf{R}(\boldsymbol{x}_{12})\,\partial_{\boldsymbol{x}_{1}}^{2}
\\
\kappa_{\m}\mathsf{R}(\boldsymbol{x}_{12})^{\dagger}\,\overset{\leftarrow}{\partial_{\boldsymbol{x}_{1}}^{2}} 
& \tilde{\mathsf{F}}(\boldsymbol{x}_{12}) \ ,
\end{bmatrix}
\end{eqnarray}
and
\begin{eqnarray}
\label{RGEQ:2pss}
\lefteqn{\hspace{-0.8cm}
\boldsymbol{\mathcal{W}}_{(\boldsymbol{\jmath},\bar{\boldsymbol{\jmath}})}^{(2)}
(\boldsymbol{x}_{1},\boldsymbol{x}_{2},t_{1},t_{2})
:=
}
\nonumber\\&&
\begin{bmatrix}
\mathcal{W}_{(\boldsymbol{\jmath},\bar{\boldsymbol{\jmath}})}^{(2,0)} & 
\mathcal{W}_{(\boldsymbol{\jmath},\bar{\boldsymbol{\jmath}})}^{(1,1)}
\\
\mathcal{W}_{(\boldsymbol{\jmath},\bar{\boldsymbol{\jmath}})}^{(1,1)\dagger} 
& \mathcal{W}_{(\boldsymbol{\jmath},\bar{\boldsymbol{\jmath}})}^{(0,2)}
\end{bmatrix}\circ(\boldsymbol{x}_{1},\boldsymbol{x}_{2},t_{1},t_{2}) \ ,
\end{eqnarray}
we can recast (\ref{RGEQ:gfflow}) into the form of an equation for the
free energy which, in compact form, reads
\begin{eqnarray}
\label{RGEQ:feflow}
\lefteqn{
\m\partial_{\m}\mathcal{W}_{\left(\boldsymbol{\jmath},\bar{\boldsymbol{\jmath}}\right)}=
}
\nonumber\\&&
\frac{1}{2}\tr\left\{(\m\partial_{\m}\boldsymbol{\mathcal{R}})\star
\left(\boldsymbol{\mathcal{W}}_{\left(\boldsymbol{\jmath},\bar{\boldsymbol{\jmath}}\right)}^{(2)}
-\boldsymbol{\mathcal{W}}_{\left(\boldsymbol{\jmath},\bar{\boldsymbol{\jmath}}\right)}^{(1)}
\boldsymbol{\mathcal{W}}_{\left(\boldsymbol{\jmath},\bar{\boldsymbol{\jmath}}\right)}^{(1)}\right)
\right\} \ .
\end{eqnarray}
Functional derivatives at zero sources of (\ref{RGEQ:feflow}) spawn a
hierarchy of equations satisfied by the full set of connected
correlation of the theory.  From (\ref{RGEQ:feflow}) we derive the
average action flow using the following two observations. First, the
very definition of Legendre transform (\ref{RGEQ:tp}) implies
\begin{eqnarray}
\label{}
\m\partial_{\m}\mathcal{W}_{\left(\boldsymbol{\jmath},\bar{\boldsymbol{\jmath}}\right)}=-
\m\partial_{\m}\mathcal{U}_{\left(\boldsymbol{u},\bar{\boldsymbol{u}}\right)} \ .
\end{eqnarray}  
Second, the evaluation of (\ref{RGEQ:2pss}) at zero sources restores 
translational invariance: 
\begin{eqnarray}
\label{RGEQ:2ps}
\boldsymbol{\mathsf{W}}^{(2)}(\boldsymbol{x}_{12},t_{12}):=\begin{bmatrix}
\mathsf{W}^{(2,0)} & \mathsf{W}^{(1,1)}
\\
\mathsf{W}^{(1,1)\dagger} & 0
\end{bmatrix}\circ(\boldsymbol{x}_{12},t_{12}) \ .
\end{eqnarray}
The matrix elements of (\ref{RGEQ:2ps}) are specified by the second
order correlation and response functions (\ref{RGEQ:cor}),
(\ref{RGEQ:res}). We may refer to them as indicators of the
``Gaussian'' part of the statistics of (\ref{intro:NS}).  We can use
(\ref{RGEQ:2ps}) and the general relation
\begin{eqnarray}
\label{}
\boldsymbol{\mathrm{I}}=\boldsymbol{\mathcal{W}}_{(\boldsymbol{\jmath},\bar{\boldsymbol{\jmath}})}^{(2)}
\star\boldsymbol{\mathcal{U}}_{(\boldsymbol{u},\bar{\boldsymbol{u}})}^{(2)} \ ,
\end{eqnarray}
following from the Legendre transform (\ref{RGEQ:tp}), to decouple the
average action into a Gaussian and an interaction part
\cite{BoDaMa94}:
\begin{eqnarray}
\label{}
\boldsymbol{\mathrm{I}}:=
\boldsymbol{\mathcal{W}}_{(\boldsymbol{\jmath},\bar{\boldsymbol{\jmath}})}^{(2)}\star
[\boldsymbol{\mathsf{W}}^{(2)-1}+\boldsymbol{\mathcal{U}}_{(\boldsymbol{u},\bar{\boldsymbol{u}})}^{(2)\,int}] \ .
\end{eqnarray}
Solving this latter relation for $\boldsymbol{\mathcal{W}}_{(\boldsymbol{\jmath},\bar{\boldsymbol{\jmath}})}^{(2)}$
\begin{eqnarray}
\label{}
\boldsymbol{\mathcal{W}}_{(\boldsymbol{\jmath},\bar{\boldsymbol{\jmath}})}^{(2)}=[\boldsymbol{\mathrm{I}}+
\boldsymbol{\mathsf{W}}^{(2)}\star\boldsymbol{\mathcal{U}}_{(\boldsymbol{u},\bar{\boldsymbol{u}})}^{(2)int}]^{-1}
\star\boldsymbol{\mathsf{W}}^{(2)} \ ,
\end{eqnarray}
allows us to finally derive the equation for the average action :
\begin{eqnarray}
\label{RGEQ:aaflow}
\lefteqn{\hspace{-0.5cm}
\m\partial_{\m}\left\{\mathcal{U}_{(\boldsymbol{u},\bar{\boldsymbol{u}})}-\frac{1}{2}[\boldsymbol{u}\,,\bar{\boldsymbol{u}}]\star 
(\m \partial_{\m}\boldsymbol{\mathcal{R}})
\star\begin{bmatrix}\boldsymbol{u}\\\bar{\boldsymbol{u}}\end{bmatrix} \right\}=
}
\nonumber\\
&&\hspace{-0.5cm}-\mathrm{tr} \,\frac{(\m \partial_{\m}\boldsymbol{\mathcal{R}})}{2}
\star\left[\sum_{n=0}^{\infty}
(-\boldsymbol{\mathsf{W}}^{(2)}\star\boldsymbol{\mathcal{U}}_{(\boldsymbol{u},\bar{\boldsymbol{u}})}^{(2)int})^{n}\right]
\star\boldsymbol{\mathsf{W}}^{(2)} \ .
\end{eqnarray}
Some observations are in order. First, the flow equation
(\ref{RGEQ:aaflow}) is effectively an equation for the reduced average
action obtained by subtracting the quadratic counter-terms associated
to the running infra-red cut-off. This is desirable because all
physical information is indeed contained in the reduced average
action.  Second, the flow in (\ref{RGEQ:aaflow}) does not depend upon
the theory under consideration which instead specify the initial
conditions for the evolution. This is a formalization of Wilson's idea
of renormalization group as a flow in the space of the probability
measures. The fixed point of the flow
does not depend on the details of the microscopic theory used as
initial condition for $\m=M$. It depends instead upon the basin of
attraction to which the initial condition belongs. Finally, solving
(\ref{RGEQ:aaflow}) exactly is equivalent to solve an infinite
non-close hierarchy of equations. Perturbative renormalization tells
us, however, that there are only a finite number of relevant coupling,
at most two for $\varepsilon\ll 1$ and $d\,\gtrsim\,2$
\cite{HoNa96,AdHoKoVa05,AdAnVa99}, determining the scaling properties
of the stochastic Navier--Stokes (\ref{intro:NS}). Based on this
observation, we now turn to the derivation of a truncation of the
right hand side of (\ref{RGEQ:aaflow}) in order to derive explicit
scaling predictions.

\section{Galilean invariance and approximation}
\label{sec:approx}

Perturbative renormalization identifies the number of relevant
couplings by diagram power counting in unit of the ultra-violet
cut-off \cite{Zinn}. Relevant couplings correspond to proper vertices
$\mathsf{U}^{(i,j)}$ proportional to powers of $M$ larger or equal
than zero. For the stochastic Navier--Stokes equations only
$\mathsf{U}^{(1,1)}$ for any $d$ and $\mathsf{U}^{(0,2)}$, for
$d\,\gtrsim\,2$ have non-negative ultra-violet degree. We can use this
information to hypothesize that (\ref{RGEQ:aaflow}) converges towards
an average action of the form
\begin{eqnarray}
\label{approx:Ansatz}
\lefteqn{
\mathcal{U}_{(\boldsymbol{u},\bar{\boldsymbol{u}})}
=\boldsymbol{u}\star\mathsf{U}^{(1,1)}\star\bar{\boldsymbol{u}}
}
\nonumber\\&&
+\frac{1}{2}\mathsf{U}^{(0,2)}(\star\bar{\boldsymbol{u}})^{2}+\frac{1}{2}(\boldsymbol{u}\star)^{2}
\mathsf{U}^{(2,1)}\star\bar{\boldsymbol{u}} \ .
\end{eqnarray}
Clearly, the Ansatz closes the hierarchy of equations spawned by
(\ref{RGEQ:aaflow}) since it is straightforward to verify that
\begin{eqnarray}
\label{}
\boldsymbol{\mathcal{U}}_{(\boldsymbol{u},\bar{\boldsymbol{u}})}^{(2)int}=
\begin{bmatrix}
\mathsf{U}^{(2,1)}\star\bar{\boldsymbol{u}} & \boldsymbol{u}\star\mathsf{U}^{(2,1)}
\\
(\boldsymbol{u}\star\mathsf{U}^{(2,1)})^{\dagger} & 0
\end{bmatrix} \ ,
\end{eqnarray}
and by (\ref{RGEQ:tp})
\begin{subequations}
\begin{eqnarray}
\label{approx:response}
\mathsf{W}^{(1,1)}=\mathsf{U}^{(1,1)\dagger\,-1} \ ,
\end{eqnarray}
\begin{eqnarray}
\label{approx:cor}
\mathsf{W}^{(2,0)}=-\mathsf{W}^{(1,1)}\star\mathsf{U}^{(0,2)}\star\mathsf{W}^{(1,1)\dagger} \ .
\end{eqnarray}
\end{subequations}
Note that
\begin{eqnarray}
\label{}
\mathcal{Z}_{\left(\boldsymbol{\jmath},\bar{\boldsymbol{\jmath}}\right)}^{(0,1)}(\boldsymbol{x},t)
=\prec\,e^{\boldsymbol{\jmath}\star\boldsymbol{v}(\boldsymbol{f}+\bar{\boldsymbol{\jmath}})}
\boldsymbol{\jmath}\star\frac{\delta\boldsymbol{v}(\boldsymbol{f}+\bar{\boldsymbol{\jmath}})}{\delta \
\bar{\boldsymbol{\jmath}}(\boldsymbol{x},t)}\,\succ \ ,
\end{eqnarray} 
implies that $\mathsf{W}^{(0,i)}=\mathsf{U}^{(i,0)}=0$ for any integer $i$.
To further evince the rationale behind (\ref{approx:Ansatz}) we observe that 
\begin{eqnarray}
\label{approx:eddy}
\lefteqn{
\check{\mathsf{U}}^{(1,1)}(\boldsymbol{p}_{1},\omega_{1}|\boldsymbol{p}_{2},\omega_{2})=
(2\,\pi)^{d+1}\,\delta^{(d)}(\sum_{i=1}^{2}\boldsymbol{p}_{i})}
\nonumber\\&&
\times \,\delta(\sum_{i=1}^{2}\omega_{i})[\imath\,\omega_{1}+\kappa\,p_{1}^{2}\,g^{(1,1)}(p_{1},\omega_{1})]
\,\mathsf{T}(\boldsymbol{p}_{1}) \ ,
\end{eqnarray}
corresponds to a ``dressing'' of the quadratic coupling in
(\ref{RGEQ:tpini}).  Differentiating with respect to $p_{1}^{2}$ at zero
wave-number and frequency the translational invariant part of
(\ref{approx:eddy}) provides a convenient non-perturbative definition
of the eddy diffusivity. We will therefore refer to
(\ref{approx:eddy}) as the ``eddy diffusivity'' vertex. Also the
``interaction'' vertex
\begin{eqnarray}
\label{approx:int1}
\lefteqn{\hspace{-0.2cm}\check{\mathsf{U}}
(\boldsymbol{p}_{1},\omega_{1},\boldsymbol{p}_{2},\omega_{2}|\boldsymbol{p}_{3},\omega_{3})=
}
\nonumber\\&&
(2\,\pi)^{d+1}\,\delta^{(d)}(\sum_{i=1}^{3}\boldsymbol{p}_{i})
\,\delta(\sum_{i=1}^{3}\omega_{i})\,\imath\,g^{(2,1)}(p_{1},\omega_{1},p_{2},\omega_{2})
\nonumber\\&&
\times\,\mathcal{P}_{(\boldsymbol{p}_{1},\boldsymbol{p}_{2})}\left\{\mathsf{T}(\boldsymbol{p}_{1})\cdot\mathsf{T}(\boldsymbol{p}_{3})
\otimes \mathsf{T}(\boldsymbol{p}_{2})\cdot\boldsymbol{p}_{3}\right\} \ ,
\end{eqnarray}
admits a similar direct interpretation from (\ref{RGEQ:tpini}). Finally, comparison with 
(\ref{RGEQ:tpini}) evinces that the ``force'' vertex
\begin{subequations}
\begin{eqnarray}
\label{approx:force}
\lefteqn{\hspace{-0.6cm}
\check{\mathsf{U}}^{(0,2)}(\boldsymbol{p}_{1},\omega_{1},\boldsymbol{p}_{2},\omega_{2})=
}
\nonumber\\&&\hspace{-0.5cm}
-(2\,\pi)^{d+1}\,\delta^{(d)}(\sum_{i=1}^{2}\boldsymbol{p}_{i})\delta(\sum_{i=1}^{2}\omega_{i})
\,g^{(0,2)}(p_{1},\omega_{1})\,\mathsf{T}(\boldsymbol{p}_{1}) \ ,
\end{eqnarray}
\begin{eqnarray}
\label{}
\lefteqn{\hspace{-0.5cm}
g^{(0,2)}(p_{1},\omega_{1}):=
}
\nonumber\\&&\hspace{-0.5cm}
\frac{1}{d-1}\sum_{i=0}^{1}
\tr\,\check{\mathsf{F}}_{(i)}(p_{1},\m)+\tilde{g}^{(0,2)}(p_{1},\omega_{1}) \ ,
\end{eqnarray}
\end{subequations}
describes (minus) the effective forcing correlation. The three vertices are,
however, not completely independent. Galilean invariance constrains
the average action to satisfy the Ward identity (see
e.g. \cite{FrTa94,To97,CoTo97} and appendix~\ref{ap:gf:ward})
\begin{eqnarray}
0=\ddot{\boldsymbol{r}}\star \boldsymbol{\bar{u}}
+\frac{\delta\,\mathcal{U}}{\delta \boldsymbol{u}}\star\left(\boldsymbol{r}\cdot\boldsymbol{\partial} \boldsymbol{u}
-\dot{\boldsymbol{r}}\right)+
\frac{\delta\,\mathcal{U}}{\delta\bar{\boldsymbol{u}}}\star \boldsymbol{r}\cdot\boldsymbol{\partial}\bar{\boldsymbol{u}} \ ,
\label{approx:effective}
\end{eqnarray}
whence it follows after standard manipulations \cite{Zinn}
\begin{eqnarray}
\label{approx:vr}
\lefteqn{
\check{\mathsf{U}}^{(2,1)}
(\boldsymbol{p}_{1},\omega_{1},\boldsymbol{0},0|\boldsymbol{p}_{3},\omega_{3})
}
\nonumber\\&&
=\boldsymbol{p}_{1}\partial_{\omega_{1}}\check{\mathsf{U}}^{(1,1)}
(\boldsymbol{p}_{1},\omega_{1}|\boldsymbol{p}_{3},\omega_{3}) \ .
\end{eqnarray}
In the context of perturbative renormalization (\ref{approx:vr}) is
used to show that if a parameter fine-tuning ensures that
$\check{\mathsf{U}}^{(1,1)}$ is finite in the limit $M$ tending to
infinity so must be $\check{\mathsf{U}}^{(2,1)}$. In general
(\ref{approx:vr}) is not sufficient to fully specify the form of the
interaction vertex in terms of $\mathsf{U}^{(1,1)}$.  If we,
furthermore, hypothesize
\begin{eqnarray}
\label{approx:int2}
g^{(2,1)}=1 \ ,
\end{eqnarray}
then (\ref{approx:vr}) implies 
\begin{eqnarray}
\label{}
g^{(1,1)}(p_{1},\omega)=g^{(1,1)}(p) \ .
\end{eqnarray}
Such an approximation is too rough to give a self-consistent model for
the full second order statistics. Our goal here is more restrictive as
it is only to derive self-consistent scaling predictions at scales
much larger than the dissipative. We therefore posit that
(\ref{approx:Ansatz}) and (\ref{approx:int2}) may serve for a
self-closure able to capture the scaling behavior of the zero frequency
sector of the theory. We also notice that a consequence of imposing
(\ref{approx:int2}), is that a generalized Taylor hypothesis
\cite{Frisch95} is verified by the two point correlation function for
which the dispersion relation
\begin{eqnarray}
\label{approx:Taylor}
\omega=\imath\,\kappa\,p^{2}\,g^{(1,1)}(p) \ ,
\end{eqnarray}
holds true.  As a final step in the derivation of our approximation we
rewrite  the  vertices  (\ref{approx:eddy}),  (\ref{approx:force})  to
decouple explicitly  the functional dependence on  the cut-off.  Thus,
we couch the eddy-diffusivity vertex into the form
\begin{eqnarray}
\label{approx:eddy_uk}
g^{(1,1)}(p;\m):=\frac{\kappa_{\m}}{\kappa}\left[\gamma^{(1,1)}\left(p;\m\right)+\check{R}\left(\frac{p}{\m}\right)
\right] \ ,
\end{eqnarray}
where now $\gamma^{(1,1)}$ is an unknown non-dimensional function which
our renormalization group equation will determine. Similarly we write
\begin{eqnarray}
\label{}
\lefteqn{
\tilde{g}^{(0,2)}(p;\m):=
}
\nonumber\\&&
\left[\lambda_{(0)}\,\m^{2-d-2\,\varepsilon}
+\lambda_{(1)}\right]\,p^{2}\,\gamma^{(0,2)}
\left(p;\m\right) \ ,
\end{eqnarray}
where we defined the Grashof numbers 
\begin{subequations}
\label{approx:Grashof}
\begin{eqnarray}
\label{approx:lambda0}
\lambda_{(0)}:=\frac{\Omega_{d}}{(2\,\pi)^{d}}\frac{F_{o}}{\kappa_{\m}^{3}\,\m^{2\,\varepsilon}} \ ,
\end{eqnarray}
\begin{eqnarray}
\label{approx:lambda1}
\lambda_{(1)}:=\frac{\Omega_{d}}{(2\,\pi)^{d}}\frac{F_{\m}}{\kappa_{\m}^{3}\,\m^{2-d}} \ ,
\end{eqnarray}
\end{subequations}
measuring the intensity of the non-local and local components of the
stochastic forcing.  In the context of perturbative renormalization
the pair (\ref{approx:Grashof}) specifies the running coupling
constant of the model \cite{HoNa96,AdHoKoVa05,AdAnVa99}.  In
(\ref{approx:Grashof}) we denoted
\begin{eqnarray}
\label{}
\Omega_{d}=\frac{2\,\pi^{d/2}}{\Gamma\left(\frac{d}{2}\right)} \ .
\end{eqnarray}

\section{Approximated renormalization group flow}
\label{sec:eqs}

The Ansatz
\begin{eqnarray}
\label{eqs:tp}
\lefteqn{
\mathcal{U}_{(\boldsymbol{u},\bar{\boldsymbol{u}})}
=\boldsymbol{u}\star\mathsf{U}^{(1,1)}\star\bar{\boldsymbol{u}}
}
\nonumber\\&&
+\frac{1}{2}\mathsf{U}^{(0,2)}(\star\bar{\boldsymbol{u}})^{2}
+(\mathsf{T}\bar{\boldsymbol{u}})\cdot [(\mathsf{T}\boldsymbol{u})\cdot\boldsymbol{\partial}](\mathsf{T}\boldsymbol{u}) \ ,
\end{eqnarray}
with $\mathsf{T}$ the transverse projector and (\ref{approx:eddy}), (\ref{approx:force}) 
specifying the Fourier representation of the order two vertices summarizes the
approximations described in the previous section. The insertion of
(\ref{eqs:tp}) into the exact renormalization group equation
(\ref{RGEQ:aaflow}) yields the equations
\begin{subequations}
\label{eqs:feqs}
\begin{eqnarray}
\label{}
\lefteqn{
\m\partial_{\m} \left\{(d-1)\,\kappa_{\m}\,p^{2}\,\gamma^{(1,1)}(p;\m)\right\}=
}
\nonumber\\&&
-\frac{1}{2}\tr \left\{\widetilde{\boldsymbol{\mathsf{W}}^{(2)}}\star 
\frac{\delta\boldsymbol{\mathcal{U}}_{(\boldsymbol{u},\bar{\boldsymbol{u}})}^{(2)int}}{\delta \bar{u}_{\alpha_{1}}}\star
\boldsymbol{\mathsf{W}}^{(2)}\star 
\frac{\delta\boldsymbol{\mathcal{U}}_{(\boldsymbol{u},\bar{\boldsymbol{u}})}^{(2)int}}{\delta u^{\alpha_{2}}}
\right\}_{\omega=0}
\nonumber\\&&
-\frac{1}{2}\tr \left\{
\boldsymbol{\mathsf{W}}^{(2)}\star 
\frac{\delta\boldsymbol{\mathcal{U}}_{(\boldsymbol{u},\bar{\boldsymbol{u}})}^{(2)int}}{\delta \bar{u}_{\alpha_{1}}}\star
\widetilde{\boldsymbol{\mathsf{W}}^{(2)}}\star 
\frac{\delta\boldsymbol{\mathcal{U}}_{(\boldsymbol{u},\bar{\boldsymbol{u}})}^{(2)int}}{\delta u^{\alpha_{2}}}
\right\}_{\omega=0} 
\end{eqnarray}
\begin{eqnarray}
\label{}
\lefteqn{
\m\partial_{\m}\left\{(d-1)\, \tilde{g}^{(0,2)}(p;\m)
\right\}=
}
\nonumber\\&&
-\frac{1}{2}\tr \left\{
\widetilde{\boldsymbol{\mathsf{W}}^{(2)}}\star 
\frac{\delta\boldsymbol{\mathcal{U}}_{(\boldsymbol{u},\bar{\boldsymbol{u}})}^{(2)int}}{\delta \bar{u}_{\alpha_{1}}}\star
\boldsymbol{\mathsf{W}}^{(2)}\star 
\frac{\delta\boldsymbol{\mathcal{U}}_{(\boldsymbol{u},\bar{\boldsymbol{u}})}^{(2)int}}{\delta \bar{u}_{\alpha_{2}}}
\right\}_{\omega=0}
\nonumber\\&&
-\frac{1}{2}\tr \left\{
\boldsymbol{\mathcal{W}}^{(2)}\star 
\frac{\delta\boldsymbol{\mathcal{U}}_{(\boldsymbol{u},\bar{\boldsymbol{u}})}^{(2)int}}{\delta \bar{u}_{\alpha_{1}}}\star
\widetilde{\boldsymbol{\mathsf{W}}^{(2)}}\star
\frac{\delta\boldsymbol{\mathcal{U}}_{(\boldsymbol{u},\bar{\boldsymbol{u}})}^{(2)int}}{\delta \bar{u}_{\alpha_{2}}}
\right\}_{\omega=0}
\end{eqnarray}
\end{subequations}
where we defined
\begin{eqnarray}
\label{}
\widetilde{\boldsymbol{\mathsf{W}}^{(2)}}:=\boldsymbol{\mathsf{W}}^{(2)}\star(\m \partial_{\m}\boldsymbol{\mathcal{R}})\star
\boldsymbol{\mathsf{W}}^{(2)} \ .
\end{eqnarray}
These equations, the explicit expression of which is given in appendix~\ref{ap:cvs}, 
admit a simple diagrammatic interpretation. Namely if we adopt the symbolic representation
\begin{subequations}
\begin{eqnarray}
\label{eqs:response}\hspace{-0.5cm}
\mathsf{W}^{(2,0)}\equiv-\mathsf{W}^{(1,1)}\star\mathsf{U}^{(0,2)}\star[\mathsf{W}^{(1,1)}]^{\dagger}
=\parbox{1.4cm}{\includegraphics[width=1.4cm]{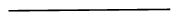}} \ ,
\end{eqnarray}
\begin{eqnarray}
\label{eqs:correlation}
\hspace{-2.5cm}\mathsf{W}^{(1,1)}\equiv[\mathsf{U}^{(1,1)}]^{\dagger-1}=
\parbox{1.4cm}{\includegraphics[width=1.4cm]{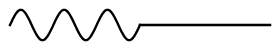}} \ ,
\end{eqnarray}
\begin{eqnarray}
\label{}
\mathsf{U}^{(2,1)}=\parbox{2.0cm}{\includegraphics[width=2.0cm]{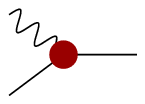}} \ ,
\end{eqnarray}
\end{subequations}
then we can couch equations (\ref{eqs:feqs}) into the form 
\begin{subequations}
\label{eqs:deqs}
\begin{eqnarray}
\label{}
\lefteqn{\hspace{-1.2cm}
\m\partial_{\m} \left\{\kappa_{\m}\,\gamma^{(1,1)}(p;\m)\right\}
}
\nonumber\\&&
=\left.\frac{1}{(d-1)\,p^{2}} \m\partial_{\m}\tr
\parbox{2.0cm}{\includegraphics[width=2.0cm]{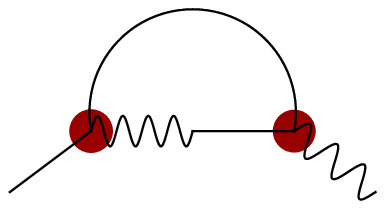}}\right|_{\omega=0} \ ,
\end{eqnarray}
\begin{eqnarray}
\label{}
\lefteqn{
\hspace{-1.0cm}
\m\partial_{\m}\left\{\left[\lambda_{(0)}\,\m^{2-d-2\,\varepsilon}
+\lambda_{(1)}\right]\,\gamma^{(0,2)}(p;\m)
\right\}
}
\nonumber\\&&
=-\left.\frac{1}{2\,(d-1)\,p^{2}}\m\partial_{\m} \tr
\parbox{2.0cm}{\includegraphics[width=2.0cm]{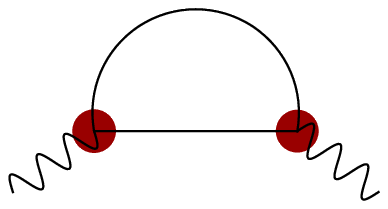}}\right|_{\omega=0} \ ,
\end{eqnarray}
\end{subequations}
if we convene to evaluate variations of response (\ref{eqs:response}) 
and correlation (\ref{eqs:correlation}) lines within the loops according 
to the rules
\begin{subequations}
\label{eqs:rules}
\begin{eqnarray}
\label{eqs:eddyrule}
\lefteqn{
\hspace{-0.4cm}
\m\partial_{\m}\kappa\,g^{(1,1)}(p;\m)\approx 
}
\nonumber\\&&
\eta_{\kappa}\,\kappa_{\m}\,\check{R}\left(\frac{p}{\m}\right)
+\kappa_{\m}\m\partial_{\m} \,\check{R}\left(\frac{p}{\m}\right) \ ,
\end{eqnarray}
\begin{eqnarray}
\label{eqs:corrule}
\lefteqn{
\hspace{-0.8cm}
\m\partial_{\m}g^{(0,2)}(p;\m)\approx \eta_{F}\,F\,_{(1)}\,\chi_{1}(p,\m)
}
\nonumber\\&&
-\sum_{i=0}^{1}F\,_{(i)}\,(\boldsymbol{p}\cdot\partial_{\boldsymbol{p}}-d_{F_{i}})
\mathcal{\chi}_{(i)}\left(p,\m\right) \ ,
\end{eqnarray}
\end{subequations}
where there appear the scaling exponents 
\begin{eqnarray}
\label{eqs:coefs}
\eta_{\kappa}:=\m\frac{d }{d \m}\ln \kappa_{\m}
\hspace{0.4cm} \& \hspace{0.4cm}
\eta_{F}:=\m\frac{d }{d \m}\ln F_{\m} \ ,
\end{eqnarray}
determined by the fixed point of the renormalization group flow and
the canonical dimensions
\begin{eqnarray}
\label{}
d_{F_{0}}=4-d-2\,\varepsilon
\hspace{1.0cm}\&\hspace{1.0cm}
d_{F_{1}}=2 \ .
\end{eqnarray}
In other words, (\ref{eqs:feqs}) imply that the functional vector field
driving the renormalization group flow with our approximation is obtained by
taking the variation of the mode coupling equations in a way adapted to
(\ref{RGEQ:Rm}). We summarize this calculation in appendix~\ref{ap:cvs}.
Here, we notice instead that after turning to non-dimensional variables 
( $\boldsymbol{p}\mapsto \boldsymbol{p}/\m$) we can rewrite (\ref{eqs:feqs})
as
\begin{subequations}
\label{eqs:eqs}
\begin{eqnarray}
\label{eqs:eddy}
\lefteqn{
[\m\partial_{\m}-\boldsymbol{p}\cdot\partial_{\boldsymbol{p}}+\eta_{\kappa}]\gamma^{(1,1)}(p)
}
\nonumber\\&&
=\eta_{F}\,G^{(1,1)}_{F}(p)-\eta_{\kappa}\,G^{(1,1)}_{\kappa}(p)-G^{(1,1)}_{o}(p) \ ,
\end{eqnarray}
\begin{eqnarray}
\label{eqs:force}
\lefteqn{
\left[\m\partial_{\m}-\boldsymbol{p}\cdot\partial_{\boldsymbol{p}}
+\tilde{\eta}_{F}\right]
\gamma^{(0,2)}(p)
}
\nonumber\\&&
=\eta_{F}\,G_{F}^{(0,2)}(p)-\eta_{\kappa}\,G_{\kappa}^{(0,2)}(p)-G_{o}^{(0,2)}(p) \ ,
\end{eqnarray}
\end{subequations}
where
\begin{eqnarray}
\label{}
\tilde{\eta}_{F}=\frac{(2-d-2\,\varepsilon)\lambda_{(0)}+\eta_{F}\,\lambda_{(1)}}
{\left(\lambda_{(0)}+\lambda_{(1)}\right)} \ .
\end{eqnarray}
The set of the $G_{k}^{(i,j)}$'s are non-linear convolutions of the
unknown functions $\gamma^{(1,1)}$, $\gamma^{(0,2)}$ with certain
integral kernels specified by the dynamics. We detail the form of
these convolutions in appendices~\ref{ap:cvs:e} and \ref{ap:cvs:f}.  
In order to fully specify the dynamics we need to associate to (\ref{eqs:eqs}) 
two renormalization conditions specifying the coefficients
(\ref{eqs:coefs}). We require

\begin{eqnarray}
\label{eqs:rc}
\gamma^{(1,1)}(p_{o})=\gamma^{(0,2)}(p_{o})=1 \ ,
\end{eqnarray}
where $p_{o}$ is the renormalization scale, i.e. the reference
infra-red scale where we suppose to measure the eddy-diffusivity and
the force amplitude.  Solving the renormalization condition
(\ref{eqs:rc}) for $\eta_{F}$, $\eta_{\kappa}$ we obtain
\begin{subequations}
\label{ets:etas}
\begin{eqnarray}
\label{rg:etak}
\lefteqn{
\eta_{\kappa}=
}
\nonumber\\&&
\hspace{-0.3cm}
\frac{G_{\star F}^{(1, 1)}\,\tilde{G}_{\star}^{(0,2)}
+\left[\frac{\lambda_{(1)}}{\lambda_{(0)}+\lambda_{(1)}}
-G_{\star F}^{(0, 2)}\right]\,\tilde{G}_{\star}^{(1,1)}}
{G_{\star F}^{(1, 1)}\,G_{\star\kappa}^{(0, 2)}+
\left[\frac{\lambda_{(1)}}{\lambda_{(0)}+\lambda_{(1)}}
-G_{\star F}^{(0, 2)}\right][1+G_{\star\kappa}^{(1, 1)}]
} \ ,
\end{eqnarray}
\begin{eqnarray}
\label{rg:etaf}
\hspace{-2.7cm}\eta_{F}=
\frac{1+G_{\star\kappa}^{(1, 1)}}{G_{\star F}^{(1,1)}}\eta_{\kappa}-\frac{\tilde{G}_{\star}^{(1,1)}}{G_{\star F}^{(1,1)}} \ ,
\end{eqnarray}
\end{subequations}
where $G^{(i,j)}_{\star\,k}\equiv G^{(i,j)}_{k}(p_{o})$ for all $i,j,k$ and
\begin{subequations}
\begin{eqnarray}
\label{}
\tilde{G}_{\star}^{(1,1)}:=(\boldsymbol{p}\cdot\partial_{\boldsymbol{p}}\gamma^{(1,1)})(\boldsymbol{p}_{o})
-G_{o}^{(1,1)}(p_{o}) \ ,
\end{eqnarray}
\begin{eqnarray}
\label{}
\lefteqn{
\hspace{-0.6cm}
\tilde{G}_{\star}^{(0,2)}:=(\boldsymbol{p}\cdot\partial_{\boldsymbol{p}}\gamma^{(0,2)})(\boldsymbol{p}_{o})
}
\nonumber\\&&
-\frac{(2-d-2\,\varepsilon)\,\lambda_{(0)}}{\lambda_{(0)}+\lambda_{(1)}}
-G_{o}^{(0, 2)}(p_{o}) \ .
\end{eqnarray}
\end{subequations}
The physical motivation behind the renormalization conditions
(\ref{eqs:rc}) is the following.  When the running cut-off $\m$ is of
the order of the ultra-violet cut-off $M$ the average action tends to
the limit (\ref{RGEQ:tp}) with forcing correlation dominated by the 
local component. In such a case we can choose
\begin{eqnarray}
\label{eqs:init}
\gamma^{(1,1)}(p)=\gamma^{(0,2)}(p)=1 \ ,
\end{eqnarray}
for any $p$: (\ref{eqs:init}) indeed specifies the initial condition for
(\ref{eqs:eqs}).  Irrespectively of $\m$, we also expect at scales
comparable with the integral scale $m^{-1}$ the bulk statistics to be
approximately Gaussian, with parameters specified by the eddy
diffusivity and the renormalized forcing amplitude. In between, as
$\m$ decreases toward $m$ we expect the onset of a non-trivial scaling
range in $\gamma^{(1,1)}$, $\gamma^{(0,2)}$ specified by the solution
of (\ref{eqs:eqs}), (\ref{ets:etas}). 
The initial value of the Grashof numbers $\lambda_{(i)}$, $i=0,1$ 
parametrize the basins of attraction of the truncated renormalization group
flow. The invariant sets of the planar dynamics
\begin{subequations}
\label{eqs:beta}
\begin{eqnarray}
\label{eqs:beta0}
\hspace{-1.3cm}
\m\partial_{\m}\lambda_{(0)}=-\lambda_{(0)}\,(3\,\eta_{k}+2\,\varepsilon) \ ,
\end{eqnarray}
\begin{eqnarray}
\label{eqs:beta1}
\m\partial_{\m}\lambda_{(1)}=-\lambda_{(1)}\,(3\,\eta_{k}+2-d-\eta_{F}) \ ,
\end{eqnarray}
\end{subequations}
characterize the possible scaling regimes that our approximations 
can capture. A priori we can distinguish four cases.

\subsection{Fixed point for $\lambda_{(0)}=\lambda_{(1)}=0$}
\label{eqs:case:A}

This is the trivial fixed point. It corresponds to decaying solutions
of the Navier--Stokes equation.

\subsection{Fixed point for $\lambda_{(0)}>0$, $\lambda_{(1)}\,\neq \,0$}
\label{eqs:case:B}

In such a case the fixed point condition is
\begin{eqnarray}
\label{etas-th}
\eta_{\kappa}=-\frac{2\,\varepsilon}{3}
\hspace{1.0cm}\&\hspace{1.0cm}
\eta_{F}=2-d-2\,\varepsilon \ ,
\end{eqnarray}
as predicted by perturbative renormalization \cite{Ho98,AdAnVa99}. 
Note that negative values of $\lambda_{1}$ are admissible if the overall 
``force'' vertex remains positive definite. If
the correlation functions also admit a limit as the integral scale $m$
tends to zero, we must observe in the scaling range
\begin{eqnarray}
\label{forfig3-a}
\gamma^{(1,1)}(p)\sim p^{-\frac{2\,\varepsilon}{3}} \ ,
\end{eqnarray} 
and
\begin{eqnarray}
\label{forfig3-b}
\gamma^{(0,2)}(p)\sim p^{2-d-2\,\varepsilon} \ .
\end{eqnarray}
We expect this behavior to be the physically correct for
$0\,<\,\varepsilon\,\ll\,1$ and $d\,>\,2$. Perturbative
renormalization in two dimensions \cite{Ho98,AdHoKoVa05} also predicts
the attainment of this fixed point.

\subsection{Fixed point for $\lambda_{(0)}>0$, $\lambda_{(1)}=0$}
\label{eqs:case:C}

The approximated renormalization group flow equations remain well
defined in the limit $\lambda_{(1)}\to 0$.  In such a case
$G^{(i,j)}_{F}(p)=0$ and (\ref{eqs:eddy}) decouples from
(\ref{eqs:force}).  Furthermore the renormalization conditions yield,
self-consistently,
\begin{eqnarray}
\label{}
\eta_{F}=0 \ .
\end{eqnarray}
In other words, the renormalization group equation has only one
relevant coupling, the eddy diffusivity. This is the situation usually
faced in perturbative renormalization under the assumption that the
spatial dimension is bounded away from two. In such a case only
$\mathsf{U}^{(1,1)}$ has non-negative ultra-violet degree. This
implies that there is no need to introduce a local counter-term in
$\mathsf{U}^{(0,2)}$ so that $F_{\m}$ is set to zero a priori. The
approximated, non-perturbative flow here devised reproduces these
features.  It is readily seen that the scaling predictions are then
the same as in case~\ref{eqs:case:B}

\subsection{Fixed point for $\lambda_{(0)}=0$, $\lambda_{(1)}\,>\,0$}
\label{eqs:case:D}

A similar fixed point, if attained, describes an energy input
dominated by its ultra-violet component independently of
$\varepsilon$. 
It is tempting to associate a similar
scenario with the $2d$ inverse cascade. The attainment of such fixed
point implies
\begin{eqnarray}
\label{}
\eta_{F}=2-d+3\,\eta_{\kappa} \ .
\end{eqnarray}
The value of $\eta_{\kappa}$ here needs to be determined dynamically.

In order to check the realizability of the aforementioned scenarios we
resorted to the numerical solution of the coupled set of equations
(\ref{eqs:eqs}), (\ref{ets:etas}) and (\ref{eqs:beta}).

\section{A simplified model}
\label{sec:toy}

Before turning to the numerical solution of (\ref{eqs:eqs}), it
is expedient to analyze a simplified version of the flow.  We therefore set
\begin{eqnarray}
\label{toy:hyp}
F_{\m}=\lambda_{(1)}=R=0 \ ,
\end{eqnarray}
and hypothesize a sharp infra-red cut-off for the power-law forcing 
\begin{eqnarray}
\label{}
F(p;\m)=H(p-\m)\,F_{o}\,p^{4-d-2\,\varepsilon} \ ,
\end{eqnarray}
where $H(x)$ is the Heaviside step function.
Since perturbative ultra-violet renormalization forbids non-local
counter-terms \cite{HoNa96,AdHoKoVa05}, these approximations are adapted only to the case
$d>2$. As a consequence, we expect (\ref{eqs:eqs}) to converge to the
fixed point of section~\ref{eqs:case:C}
\begin{subequations}
\label{toy:eqs}
\begin{eqnarray}
\label{toy:eddy}
\left(\boldsymbol{p}\cdot\partial_{\boldsymbol{p}}+\frac{2\,\varepsilon}{3}\right)\gamma_{\star}^{(1,1)}(p)
=\frac{G^{(1,1)}_{o}(p)}{\lambda_{(0)}} \ ,
\end{eqnarray}
\begin{eqnarray}
\label{toy:force}
\left[\boldsymbol{p}\cdot\partial_{\boldsymbol{p}}
-(2-d-\varepsilon)\right]
\gamma_{\star}^{(0,2)}(p)
=\frac{G_{o}^{(0,2)}(p)}{\lambda_{(0)}} \ ,
\end{eqnarray}
\end{subequations}
with $G_{o}^{(1,1)}$, $G_{o}^{(0,2)}$ respectively specified by
\begin{eqnarray}
\label{toy:eddydrift}
\lefteqn{\hspace{-0.2cm}
\frac{G_{o}^{(1,1)}(p)}{\lambda_{(0)}}=\frac{C_{d}}{2\,p^{2}}\int_{-1}^{1}d\phi\,
\frac{(1-\phi^{2})^{\frac{d-1}{2}}}{P^{2}} \times}
\nonumber\\&&\hspace{-0.3cm}
\frac{\left[(d-1)\,p^{3}(p- 2\,\phi)+(d-3)\, p^{2}\,
+2\, \phi\,p\,\right]}{g^{(1,1)}(1)\,[g^{(1,1)}(1)+P^{2}
\,g^{(1,1)}(P)]} \ ,
\end{eqnarray}
and
\begin{eqnarray}
\label{}
\label{toy:forcedrift}
\lefteqn{\hspace{-0.4cm}
\frac{G_{o}^{(0,2)}(p)}{\lambda_{(0)}}=
\frac{C_{d}}{2}\int_{-1}^{1}d\phi\,
\frac{(1-\phi^{2})^{\frac{d-1}{2}}\,g^{(0,2)}(P)}{ P^{4}} \times}
\nonumber\\&&
\frac{[(d-1)\,p^{2}-2\, d\,p\,k\,\phi+2\,k^2 \left(d+2\, \phi^{2}-2\right)]}
{g^{(1,1)}(1)\,g^{(1,1)}(P)\,[g^{(1,1)}(1)+P^{2}\,g^{(1,1)}(P)]} \ .
\end{eqnarray}
In (\ref{toy:eddydrift}), (\ref{toy:forcedrift}) we used the notation
\begin{eqnarray}
\label{}
P:=\sqrt{1+p^{2}+2\,\phi\,p} \ .
\end{eqnarray}
In the limit $p\,\gg\,1$ we can approximate (\ref{toy:eddy}) as
\begin{eqnarray}
\label{qual:11}
\left(p\partial_{p}+\frac{2\varepsilon}{3}\right)\gamma_{\star}^{(1,1)}(p)\approx
\frac{(d-1)}{2\,d\,p^{2}\,\gamma_{\star}^{(1,1)}(p)
\gamma_{\star}^{(1,1)}(1)} \ ,
\end{eqnarray}
whence we infer the leading scaling behavior
\begin{eqnarray}
\label{toy11}
\gamma^{(1,1)}_{\star}(p)\overset{p\uparrow \infty}{\sim}\left\{
\begin{array}{ll}
p^{-\frac{2\,\varepsilon}{3}} \hspace{0.2cm}&\hspace{0.2cm} 0\,<\,\varepsilon\,<\,\frac{3}{2}
\\[0.2cm]
p^{-1} \hspace{0.2cm}&\hspace{0.2cm} \frac{3}{2}\,<\,\varepsilon\,
\end{array}\right. \ ,
\end{eqnarray}
under the self-consistence condition
\begin{eqnarray}
\label{toy:largep}
p\,\gg\,\left|\frac{(d-1)}{2\,d\,c_{+}
\gamma_{\star}^{(1,1)}(1)\,\left(1-\frac{2\,\varepsilon}{3}\right)}\right|^{\frac{3}{6-4\,\varepsilon}}\,\gg\,1 \ .
\end{eqnarray}
Logarithmic corrections may be possible at $\varepsilon=3/2$.
Similarly, we can approximate (\ref{toy:force}) as
\begin{eqnarray}
\label{}
\lefteqn{
\left(p\partial_{p}+d+2\,\varepsilon-2\right)\gamma^{(0,2)}_{\star}(p)\approx
}
\nonumber\\&&
\frac{p^{2-d-2\,\varepsilon}+\gamma^{(0,2)}_{\star}(p)}
{\gamma^{(1,1)}_{\star}(p)}\left(p\partial_{p}+\frac{2\varepsilon}{3}\right)\gamma_{\star}^{(1,1)}(p) \ ,
\end{eqnarray}
in the non-dimensional wave-number range defined by (\ref{toy:largep}). 
The corresponding scaling prediction is 
\begin{eqnarray}
\label{toy02}
\gamma^{(0,2)}_{\star}(p)\overset{p\,\gg\,1}{\sim}\left\{
\begin{array}{ll}
p^{2-d-2\varepsilon} \hspace{0.2cm}&\hspace{0.2cm} 0\,<\,\varepsilon\,<\,\frac{3}{2}
\\[0.2cm]
p^{2-d-2\varepsilon+\left(\frac{2\,\varepsilon}{3}-1\right)} \hspace{0.2cm}&\hspace{0.2cm} \frac{3}{2}\,<\,\varepsilon\,
\end{array}
\right. \ .
\end{eqnarray}
The conclusion is that the model problem kinetic energy
spectrum should scale in agreement with the prediction of the
perturbative renormalization group:
\begin{eqnarray}
\label{toyE}
\mathcal{E}\left(p\right)\sim p^{d-1}\frac{p^{2-d-2\,\varepsilon}+\gamma^{(0,2)}_{\star}(p)}{\gamma^{(1,1)}_{\star}(p)}
\sim p^{1-\frac{4\varepsilon}{3}} \ .
\end{eqnarray}

 \begin{figure}[!t]
  \centerline{\includegraphics*[width=0.9\linewidth]{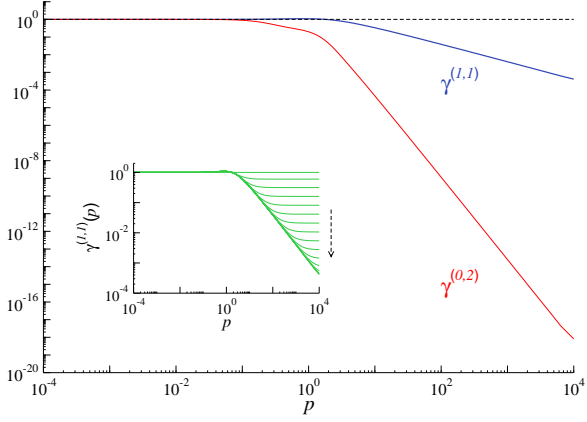}}
  \caption{(Color online)  Result of the numerical  integration of the
    dimensionless   renormalized  functions   $\gamma^{(1,1)}(p)$  and
    $\gamma^{(0,2)}(p)$ for $d=3$ and $\varepsilon=2$. The dashed line
    indicates    the    value     $1$.     Inset:    convergence    of
    $\gamma^{(1,1)}(p)$     from    its    initial     (unforced limit) value
    $\gamma^{(1,1)}(p)=1$ toward stationarity (as indicated by the arrow).}
  \label{fig:example}
\end{figure}

The eddy diffusivity and the force vertices, however, individually
deviate from the perturbative renormalization group prediction. In
particular the eddy diffusivity as observed first in \cite{MoWe95}
saturates to an $\varepsilon$ independent value for
$\varepsilon\,>\,3/2$. In Fig~\ref{fig:onerenorm}  we show that the above predictions 
compare favorably with the numerical integration of (\ref{toy:eqs}).  For
$0\,<\,\varepsilon\,<\,2$, these results are also consistent with the
direct numerical simulations of \cite{SaMaPa98,BiCeLaSbTo04}.

\section{Numerics}
\label{sec:numerics}

\begin{figure}[!t]
  \centerline{\includegraphics*[width=0.9\linewidth]{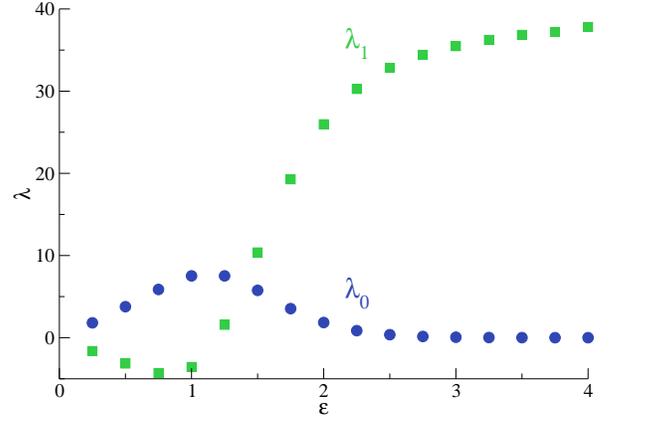}}
  \caption{(Color online) Dependence of the  fixed  point
    $(\lambda_0,\lambda_1)$  (blue dots)  on $\varepsilon$  and $d=3$.
    The  fixed point tends  toward $(0,0)$  as $\varepsilon\rightarrow
    0$.}
  \label{fig:fp3d}
\end{figure}

We integrated numerically the set of equations (\ref{eqs:eqs}) for the
eddy diffusivity and the renormalized forcing amplitude, and equations
\eref{eqs:beta}  for   the  coupling  constants,   together  with  the
renormalization conditions \eref{ets:etas}.
  
We proceeded by discretizing the  momentum space on a logarithmic mesh
for $p$  and a linear  mesh for $\phi$.  The domain of  $p$ considered
extends  from $10^{-4}$  to $10^4$  and was  covered by  $200$ points,
corresponding to a logarithmic spacing of $\approx 0.092$. The angular
domain $[-1,1]$  was covered by  $100$ points.  Moreover, in  order to
improve  accuracy   in  the  scaling  range   we  modeled  wave-number
logarithmic derivatives $\boldsymbol{p}\cdot\partial_{\boldsymbol{p}}$
using   $5$-point   finite-difference   expressions.  For   the   mass
differential  $m_\mathrm{r}\partial_{m_\mathrm{r}}$ we,  instead, used
$2$-point finite-difference expressions.  This mesh was fine enough to
observe good continuous convergence of the flow equations.

We integrated the flow equations with initial conditions
(\ref{eqs:init}), over $\m$ from $10$ down to $10^{-9}$, using a
simple Euler explicit method with logarithmic integration steps.
Finally, we estimated integrals using a trapezoidal rule on the linear
and the logarithmic mesh. The initial values of the Grashof numbers
$\lambda_{(0)}$ and $\lambda_{(1)}$ where randomly sampled in the
domain $0.01 < \lambda_{(i)} < 10$. The non-local force $\chi_{(0)}$
appearing in equation (\ref{NLforce}), consenquently in the
convolutions of equations (\ref{conv-V11}) and (\ref{conv-V02}) was
chosen as
\begin{equation} \label{chi0}
\chi_{(0)}(p) = \frac{p^2}{(p^2 + \mu_0^2)^{(d-2+2\varepsilon)/2}} \ ,
\end{equation}
with $\mu_0 = 0.1$.

As an example of the numerical integration scheme that we have used we
show in Fig.~\ref{fig:example} the results for $\gamma^{(1,1)}(p)$ and
$\gamma^{(0,2)}(p)$  for $d=3$  and  $\varepsilon=2$.  Both  functions
satisfy smoothly  the renormalization condition at  the infrared limit
$p_o=10^{-4}$ while exhibiting a  power-law decay in the ultra-violet.
We observed  the same qualitative  behavior for any values  of $d=2,3$
and     $0\,<\,\varepsilon\,\leq\,4$.      In     the     inset     of
Fig.~\ref{fig:example}  we show  the regular  convergence of  the eddy
diffusivity toward is final value.

We first discuss our numerical  results in three dimensions.

\begin{figure}[!t]
\psfrag{LE}[b][b][1.2]{$\Lambda^{(\mathcal{E})}$}
\psfrag{eps}[b][b][1.2]{$\varepsilon$}
{\includegraphics*[width=0.9\linewidth]{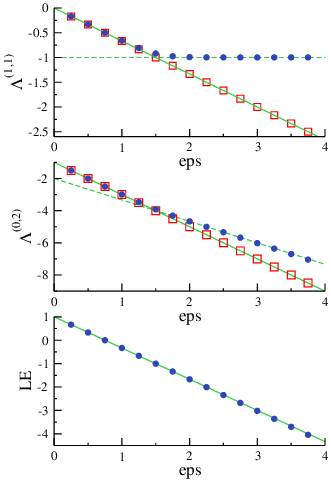}
}
  \caption{(Color    online)    Scaling    exponents   $\Lambda$    in
    Eq.~\ref{def:Lambda}  of the dimensionless  renormalized functions
    $\gamma^{(1,1)}(p)$  and  $\gamma^{(0,2)}(p)$  and of  the  energy
    spectrum $\mathcal{E}$  as a function of $\varepsilon$  (blue solid circles)
    and $d=3$. We also show in  red open squares the dependence of the
    scaling exponents $\eta_\kappa$ (upper panel) and $\eta_F$ (middle
    panel)  on  $\varepsilon$.   The  solid lines  correspond  to  the
    respective renormalization group scaling  of equations
    \eref{etas-th},   and  to   $p^{-4\varepsilon/3}$  for   the  energy
      spectrum.  The dashed lines in the upper and middle panels stand
      for   the   scalings   $p^{-1}$   and   $p^{1-d-4\varepsilon/3}$
      respectively.}
  \label{fig:scaling3d}
\end{figure}

\subsection{3d}
\label{sec:3d}

For  each  fixed  value  of  $\varepsilon$ in  $(0,4]$,  we  used  the
numerical  scheme   described  above  to   integrate  the  equations
(\ref{eqs:eqs})  and obtained, for  any initial  value of  the Grashof
numbers for which we  found convergence, a single stationary solution.
This means that for each value of $\varepsilon$ there exist one single
fixed     point      $(\lambda_{(0)},\lambda_{(1)})$     only.      In
Fig.~\ref{fig:fp3d},  we show  the dependence  of the  fixed  point on
$\varepsilon$.   We  have  noted  a  slower  convergence  towards  the
solution  as $\varepsilon\rightarrow  0$, making  hard to  explore the
perturbative  regime $\varepsilon \ll  1$.  Nevertheless,  our results
suggest that  the trivial fixed  point of section  \ref{eqs:case:A} is
reached in the limit of vanishing $\varepsilon$.

Surprisingly, for small values of $\varepsilon$, $\lambda_{(1)}<0$ and
becomes positive for a value of $\varepsilon$ between $1$ and
$1.25$.   For
$\varepsilon>2 $, $\lambda_{(0)}$ decreases exponentially, but we
always find a positive value.

To  determine   the  ultra-violet  scaling   law  as  a   function  of
$\varepsilon$, we computed
\begin{equation} 
\label{def:Lambda}
\Lambda^{(x,y)} \equiv \lim_{p\rightarrow\infty} \frac{\log \gamma^{(x,y)}}
{\log p} \ ,
\end{equation}
for $(x,y)=(1,1)$  or $(0,2)$, which  defines the scaling  exponent of
the respective  function. We denote with  $\Lambda^{(\mathcal{E})}$ the analogous
measure for the energy spectrum.

In   Fig.~\ref{fig:scaling3d}   we    show   the   scaling   exponents
$\eta_\kappa$ (red open squares in  the upper panel) and $\eta_F$ (red
open squares in the middle  panel) as a function of $\varepsilon$. Our
numerical  results are  in  excellent agreement  with the  theoretical
predictions  \eref{etas-th} (solid  lines), meaning  that  our closure
yields the perturbative renormalization scaling. In the same figure we
also  show  the scaling  exponent  of  the dimensionless  renormalized
functions  $\gamma^{(1.1)}$  (upper  panel), $\gamma^{(0,2)}$  (middle
panel)  and of the  energy spectrum  (lower panel),  as a  function of
$\varepsilon$.

\begin{figure}[!t]
  \centerline{\includegraphics*[width=0.9\linewidth]{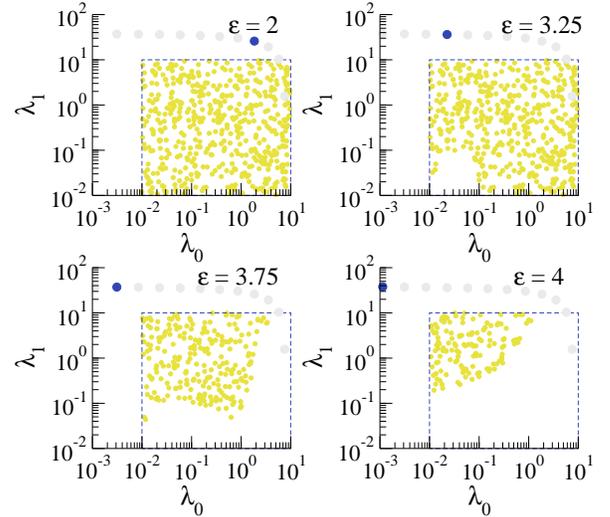}}
  \caption{(Color online)  Basin of attraction  of the fixed  point in
    three  dimensions, for  different values  of  $\varepsilon$.  Each
    yellow  (light  grey) dot,  stands  for  an  initial condition  of
    $(\lambda_{(0)},\lambda_{(1)})$ for which convergence was reached.
    The square with dashed sides  indicates the domain in which random
    initial conditions were drawn.  The light grey solid circles stand
    for    the    trajectory   of    the    fixed    point   in    the
    $\lambda_{(0)}$-$\lambda_{(1)}$  plane and  the  blue (dark  grey)
    circle,   to  the   fixed  point   for  the   specific   value  of
    $\varepsilon$.}
  \label{fig:basin3d}
\end{figure}

We  observe   two  different  regimes.   In  the   first  regime,  for
$\varepsilon<3/2$,  the  eddy diffusivity  and  the forcing  amplitude
scale in agreement with  perturbative renormalization, as obtained
in    \eref{forfig3-a}    and    \eref{forfig3-b}.     Instead,    for
$\varepsilon>3/2$,   both  fields   deviate   individually  from the
perturbative renormalization prediction.  In particular,  in this regime the eddy
diffusivity  scales as  $\gamma^{(1,1)}\sim  p^{-1}$ independently  of
$\varepsilon$.    This  saturation   has  been   predicted   first  in
\cite{MoWe95}.   More  interestingly,  the  deviation of  the  forcing
amplitude is  such that  the energy spectrum  scaling is  in agreement
with  perturbative  renormalization  {\it i.e.},  $\mathcal{E}\sim
p^{1-4\varepsilon/3}$, for all  $\varepsilon$. 
Moreover, the deviations of  the   eddy  diffusivity  and   the  forcing  
amplitude   from  the perturbative  renormalization  coincide with  those  
predicted by  our simplified  model,  equations   \eref{toy11}  and  \eref{toy02}.  

\begin{figure}[!t]
\psfrag{LE}[b][b][1.2]{$\Lambda^{(\mathcal{E})}$}
\psfrag{eps}[b][b][1.2]{$\varepsilon$}
{\includegraphics*[width=0.9\linewidth]{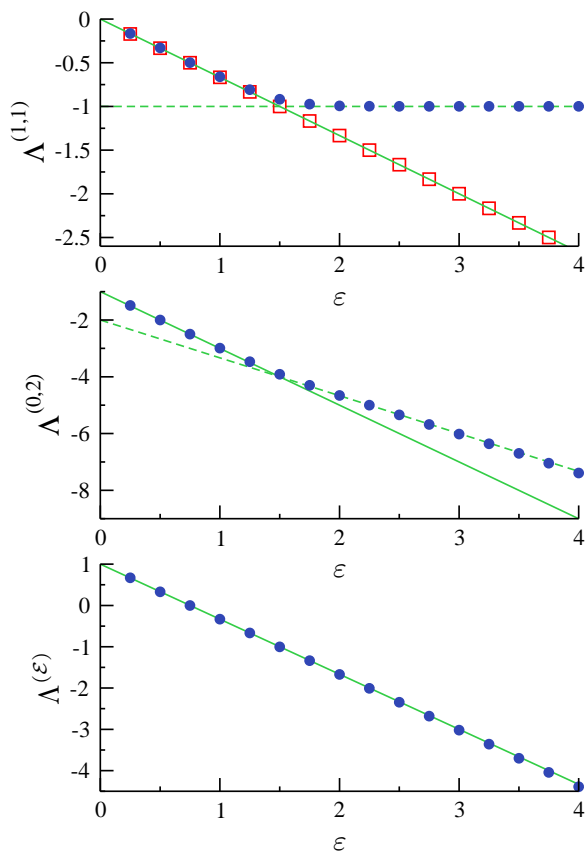}
}
\caption{(Color online) Scaling exponents $\Lambda$ in
  Eq.~\ref{def:Lambda} of the dimensionless renormalized functions
  $\gamma_\star^{(1,1)}(p)$ and $\gamma_\star^{(0,2)}(p)$ and of the
  energy spectrum $\mathcal{E}$ as a function of $\varepsilon$ (blue
  solid circles) for the simplified model of section~\ref{sec:toy} and
  $d=3$.  The red open squares in the upper panel correspond to
  $\eta_\kappa$.  The solid and dashed lines correspond to the
  predicted scaling of equations \eref{toy11}, \eref{toy02} and
  \eref{toyE}, for $\varepsilon < 3/2$ and $\varepsilon>3/2$
  respectively.}
  \label{fig:onerenorm}
\end{figure}

Finally, we would like to remark some properties of the convergence of
the numerical  scheme that we have  used.  As we  mentioned above, the
initial seed for the integration scheme comprises the initial value of
the Grashof  numbers. We have  chosen this initial numbers  by drawing
$\lambda_{(0)}$  and $\lambda_{(1)}$  as random  values in  the domain
$[0.01,10]$.   By  doing this,  we  found  that  the solution  of  our
numerical   scheme  always   converged   to  the   fixed  point   when
$\varepsilon<3$.  However,  for larger $\varepsilon$,  we noticed that
this was no  longer the case.  For $\varepsilon>3$  some of the initial
conditions   failed    to   converge.     This   can   be    seen   in
Fig.~\ref{fig:basin3d} in  which we show as yellow  (light grey) dots,
those initial conditions that converged  to the fixed point. We notice
that     the     basin    of     attraction,     limited    to     the
$[0.01,10]\times[0.01,10]$    domain,    shrinks   as    $\varepsilon$
grows. While we have no ultimate explanation  for this behavior, it may be due
 either  to the  very small values  that $\lambda_{(0)}$  attain for
$\varepsilon>3$  or, more trivially,  to the  fact that  our numerical
scheme fails to  converge to the fixed point (shown  as the blue (dark
grey) circle), when the initial condition is too far from it.

\subsection{Single renormalization condition}

We have solved the simplified model of section~\ref{sec:toy} simply by
setting $\eta_F=0$ and using  the numerical scheme described above, by
integrating    equations    (\ref{eqs:eqs}),   \eref{eqs:beta0}    and
\eref{rg:etak}.   In  Fig.~\eref{fig:onerenorm} we  show  the results  that
corroborate  the   predicted  behavior  of   equations  \eref{toy11},
\eref{toy02} and \eref{toyE}.

In summary, we have obtained that the stationary solution to equations
(\ref{eqs:eqs}) is  described by equations  \eref{toy11}, \eref{toy02}
and \eref{toyE}, irrespectively if we  impose the system to either one
or two renormalization conditions.

\begin{figure}[!t]
  \centerline{\includegraphics*[width=0.9\linewidth]{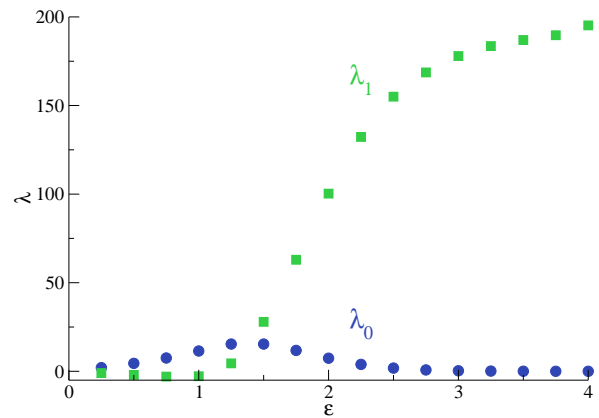}}
  \caption{(Color online) Dependence of the  fixed  point
    $(\lambda_0,\lambda_1)$  (blue dots)  on $\varepsilon$  and $d=2$.
    The  fixed point tends  toward $(0,0)$  as $\varepsilon\rightarrow
    0$.}
  \label{fig:fp2d}
\end{figure}

\subsection{2d}
\label{sec:2d}

In two dimensions the  results are in perfect
agreement with the predictions of equations \eref{toy11}, \eref{toy02}
and \eref{toyE}, meaning that the fixed point found is consistent with
the perturbative  renormalization prediction. To start the discussion  we show in
Fig.~\ref{fig:fp2d}    the   fixed   point    for   several    values   of
$\varepsilon$. The behavior  of the fixed point in  two dimensions is
qualitatively the same as in  three dimensions, namely the fixed point
$(\lambda_0,\lambda_1)$  tends to  $(0,0)$ as  $\varepsilon$  tends to
zero;  for  $\varepsilon\lessapprox 1$,  $\lambda_{(1)}<0$ and  becomes
positive  for a  value of  $\varepsilon$ between  $1$ and  $1.25$;  for
$\varepsilon>2 $, $\lambda_{(0)}$ decreases exponentially.

In  Fig.~\ref{scaling2d} we  show the  scaling  exponent $\eta_\kappa$
(red open squares in the  upper panel) as a function of $\varepsilon$,
in  agreement with  the prediction  \eref{etas-th}. Moreover,  we also
show the scaling exponent  of the dimensionless renormalized functions
$\gamma^{(1.1)}$ (upper panel), $\gamma^{(0,2)}$ (middle panel) and of
the energy spectrum (lower panel), exhibiting the same behavior as in
three dimensions, described by equations \eref{toy11}, \eref{toy02}
and \eref{toyE}.

\begin{figure}[!t]
  \centerline{\includegraphics*[width=0.9\linewidth]{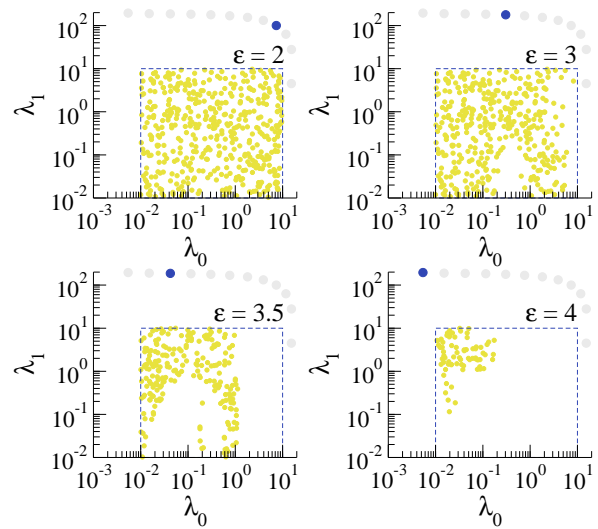}}
  \caption{(Color  online)  Basin of attraction  of the fixed  point in
    two  dimensions, for  different values  of  $\varepsilon$.  Each
    yellow  (light  grey) dot,  stands  for  an  initial condition  of
    $(\lambda_{(0)},\lambda_{(1)})$ for which convergence was reached.
    The square with dashed sides  indicates the domain in which random
    initial conditions were drawn.  The light grey solid circles stand
    for    the    trajectory   of    the    fixed    point   in    the
    $\lambda_{(0)}$-$\lambda_{(1)}$  plane and  the  blue (dark  grey)
    circle,   to  the   fixed  point   for  the   specific   value  of
    $\varepsilon$.}
  \label{fig:basin2d}
\end{figure}

Finally, as it was the case  in three dimensions, in two dimensions we
also   observed   that   the   basin   of   attraction   shrinks   for
$\varepsilon \gtrapprox  3$, as is seen in Fig.~\ref{fig:basin2d}.

\begin{figure}[!t]
\psfrag{LE}[b][b][1.2]{$\Lambda^{(\mathcal{E})}$}
\psfrag{eps}[b][b][1.2]{$\varepsilon$}
{\includegraphics*[width=0.9\linewidth]{scaling-2d.eps}}
\caption{(Color online) Scaling    exponents   $\Lambda$    in
    Eq.~\ref{def:Lambda}  of the dimensionless  renormalized functions
    $\gamma^{(1,1)}(p)$  and  $\gamma^{(0,2)}(p)$  and of  the  energy
    spectrum $\mathcal{E}$  as a function of $\varepsilon$  (blue solid circles)
    and $d=2$. We also show in  red open squares the dependence of the
    scaling exponents $\eta_\kappa$ (upper panel) and $\eta_F$ (middle
    panel)  on  $\varepsilon$.   The  solid and dashed lines correspond
  to the predicted scaling of equations \eref{toy11}, \eref{toy02} and
  \eref{toyE},   for  $\varepsilon   <   3/2$  and   $\varepsilon>3/2$
  respectively.
}
  \label{scaling2d}
\end{figure}

\section{Conclusions}
\label{sec:concl}

Power-law forcing provides us with a control parameter, $\varepsilon$,
continuously changing the energy input from ultra-violet, as if due to
thermal stirring, to infra-red as it is needed to interpret the
stochastic Navier--Stokes as a model of fully developed Newtonian
turbulence. The limit of vanishing $\varepsilon$ can be systematically
investigated using the general principles of perturbative ultra-violet
renormalization. These principles yield in three spatial dimensions
the expression of the critical, fixed point, theory for vanishing
$\varepsilon$.  For fully developed turbulence the critical theory is
not known, only some extrapolations can be made from the perturbative
limit.  The validity of these extrapolations is an open important
question since they are based on the assumptions of the absence of any
non-perturbative renormalization group fixed point and, provided this
assumption holds, require controlling the limit of infinite integral
scale of any statistical indicator of the theory after their
perturbative expressions are re-summed for finite $\varepsilon$. The
inquire of the Kraichnan model passive advection (see
e.g. \cite{FaGaVe01} and references therein for review) has in recent
years shed much light on how the limit of infinite integral scale can
be investigated in a field theory model of fully developed
turbulence. Namely, in the context of the Kraichnan model ultra-violet
renormalization reduces to a trivial operation whilst the scaling
properties of relevant physical indicators such as structure functions
are fully specified by the analysis of composite operators (see
e.g. \cite{AdAnVa98} and result discussion in \cite{KuMG07}).

In this paper, we devise the simplest possible model of
non-perturbative renormalization group flow complying with the
requirements imposed by the general principles of ultra-violet
renormalization as well as verifying the symmetries enjoyed by the
stochastic Navier--Stokes equation. Specifically, these requirements
translate in two classes of constraints. Vertices of the effective
action must satisfy the Ward identities stemming from Galilean
symmetry and space translational invariance. Furthermore, we adhere to
the postulate of ultra-violet renormalization that no counter-term,
can be consistently associated to non-local coupling. In other words,
no independent renormalization constant can be associated either to
the non-local forcing or to pressure. It is worth repeating here that
explicit check show that non-local renormalization conditions yield
inconsistencies already at second order in the perturbative expansion
in powers of $\varepsilon$ (see e.g. \cite{AdHoKoVa05}).

The intrinsic limitation of state-of-the art non-perturbative
renormalization methods is that it allows us to derive explicit
expressions only if we take into account a finite number of vertices
in the renormalization group flow. As a guideline to operate this
otherwise unjustified truncation, we restrict ourselves to
interactions which can be assessed as relevant under renormalization
at perturbative level. This is of course a dramatic approximation. We
were encouraged in taking this step by the results, to some extent
surprising, of \cite{CaChDeWs09} where it was shown that similar
approximations appear to be able to capture the existence of a
non-perturbative fixed point for the Kardar-Parisi-Zhang stochastic
partial differential equation. This latter model shares with the
stochastic Navier--Stokes equation invariance under Galilean
transformations and convergence towards a non-Boltzmann steady state.
An important difference between these two models resides, however, in
the non-locality of the interactions that the incompressibility
condition brings forth for Navier--Stokes. In our average action
Ansatz (\ref{approx:Ansatz}) incompressibility simply appears in the
form of transversal projectors acting on the classical field.  In
spite of this simple expression, the consequences of incompressibility
are evident. The non-perturbative fixed point of the
Kardar-Parisi-Zhang equation is suppressed. We also observe saturation
to an $\varepsilon$-independent value of the scaling dimension of the
eddy diffusivity at $\varepsilon=3/2$.  Perturbative renormalization
attributes to any integer power $n$ of the velocity field the scaling
dimension $n\,(1-2\,\varepsilon/3)$. This means that the saturation we
observe occurs exactly at the value of $\varepsilon$ when the velocity
field (as well as all its integer powers) becomes an infra-red
relevant operator.  The fact may well be the indication of a change of
critical behavior towards a regime not captured by our truncation. We
do not observe saturation for $\varepsilon\,>\,2$ of the energy
spectrum to the Kolmogorov value $-5/3$ for $d=3$, neither the inverse
cascade $-5/3$ energy spectrum for $0\,<\,\varepsilon\,<\,2$ and
$d=2$.  If we identify the universality of the $-5/3$ energy spectrum
in the above $\varepsilon$ domain with the presence of a scaling
regime characterized by a constant energy flux, the inference is that
it is not possible to describe a constant flux scaling regime in terms
of an effective action comprising the vertices relevant under
renormalization at perturbative level. Conversely, the average action
Ansatz (\ref{approx:Ansatz}) yields scaling predictions in agreement
with direct numerical simulations whenever the energy input at
phenomenological level is not expected to sustain a constant flux
solution of the Navier-Stokes equation ($0\,<\,\varepsilon\,<\,2$ for
$d=3$ and $2\,<\,\varepsilon\,<\,3$ in $d=2$). Phenomenological
reasoning suggests (see discussion in \cite{Frisch95,EyGo94}) that the
scaling properties of the constant flux solution are the consequence
of the ``\emph{localness}'' of the interactions  within
the turbulent fluid.  This means that after isolating transport,
``sweeping'', terms the critical theory should be described only by
couplings involving local interactions in wave-number space. If this
phenomenological reasoning is correct, constructing a renormalization
group flow in the universality class of the constant flux solution
poses a severe difficulty. On the one hand, our present results
indicate that the flow should encompass in the Ansatz average action
at least the set of proper vertices contributing to the flux.  On the
other, it is not a-priori evident how to reconcile these coupling with
the requirement of localness.

As a conclusive remark we observe that renormalization methods 
may also have spin-offs for engineering applications. 
Obtaining, for example, a priori estimates for the eddy
diffusivity and the Kolmogorov constant is very important for devising
reliable large eddy simulations of turbulent flows \cite{Sagaut}.
In \cite{YaOr86} it was suggested that renormalized
perturbation theory could be used to obtain quantitative predictions
for the Kolmogorov constant. Whilst the treatment of the problem in
\cite{YaOr86} can only be considered phenomenologically correct (see
discussion in \cite{Ey94b} and especially in section 2.10 of
\cite{AdAnVa99}), a controlled calculation of the Kolmogorov constant
up to $O(\varepsilon^{3})$ in the renormalized perturbation theory
evaluated for $\varepsilon=2$ in the limit of large spatial dimension
can be found in \cite{AdAnGoKiKo08}. The result
$C_{K}\sim 1.5+O(\varepsilon^{3},1/d)$ of \cite{AdAnGoKiKo08} 
is in reasonable agreement with experimental and numerical measurements
\cite{Sre95,YeZh97}. The non-perturbative renormalization flow devised
in this paper cannot be used in the present form to give predictions
for indicators beyond scaling exponent. The reason is that the finite 
renormalization conditions we imposed only fix the ratio
$F_{o}/\kappa^{3}$ between the ``bare'' parameters of the stochastic Navier--Stokes 
equation. In other words, we did not specify (neither had the need of specifying) 
the units in which the energy input is measured. Such way of proceeding is perfectly 
in line with the general renormalization group ideology which aims
at determining scaling exponents as only indicators of universality 
classes. It is possible, however, to envisage imposing different renormalization 
conditions fully specifying the values of the ``bare'' parameters $F_{o}$, and $\kappa$. 
This is an issue which we leave for future work.

\section{Acknowledgments}

We are grateful to Luca Peliti for pointing out to us references
\cite{LiFi86,LiFi87} and their potential relevance for a
renormalization group theory for the $2d$ inverse cascade. The work
PMG was supported by Finnish Academy CoE “Analysis and Dynamics” and
from the KITP program ``The nature of Turbulence'' (grant No. NSF
PHY05-51164). The authors acknowledge support from the ESF and
hospitality of NORDITA where part of this work has been done during
their stay within the framework of the "Non-equilibrium Statistical
Mechanics" program.

\appendix

\section{Variations of the generating of function}
\label{ap:gf}

\subsection{Renormalization group flow}
\label{ap:gf:rg}

Let us consider the deformation of (\ref{intro:NS}) induced by the
replacements $\kappa\mapsto\kappa+\kappa_{\m}R$ and
$\boldsymbol{f}\mapsto
\boldsymbol{f}^{\prime}+\bar{\boldsymbol{\jmath}}$.  We suppose that
$\boldsymbol{f}^{\prime}$ is obtained from applying to
$\boldsymbol{f}$ an high pass filter with infra-red cut off $\m$. We
have then
\begin{eqnarray}
\label{ap:gf:v}
\m\partial_{\m}\mathcal{Z}_{\left(\boldsymbol{\jmath},\bar{\boldsymbol{\jmath}}\right)}
=\prec\,e^{\boldsymbol{\jmath}\star\boldsymbol{v}}\boldsymbol{\jmath}\star(\m\partial_{\m}\boldsymbol{v})\,\succ \ ,
\end{eqnarray}
with
\begin{eqnarray}
\label{ap:gf:rg:var}
\lefteqn{\hspace{-1.0cm}
\m\partial_{\m}\boldsymbol{v}(\boldsymbol{x},t;\bar{\boldsymbol{\jmath}}+\boldsymbol{f})
=\frac{\delta \boldsymbol{v}(\boldsymbol{x},t;\bar{\boldsymbol{\jmath}}+\boldsymbol{f}
)}{\delta \bar{\boldsymbol{\jmath}}}
\star}
\nonumber\\&&
\hspace{-0.3cm}
\left\{(\m\partial_{\m}\kappa_{\m}\,R)\star\partial^{2}\boldsymbol{v}+(\m\partial_{\m}\boldsymbol{f}^{\prime})\right\} \ .
\end{eqnarray}
In (\ref{ap:gf:rg:var}) the fluctuating response 
function satisfies
\begin{eqnarray}
\label{ap:gf:causal}
\frac{\delta \boldsymbol{v}(\boldsymbol{x}_{1},t_{1}
)}{\delta \bar{\boldsymbol{\jmath}}(\boldsymbol{x}_{2},t_{2})}
=0\hspace{0.5cm} \forall\,t_{2}\leq t_{1} \ .
\end{eqnarray}
We furthermore interpret the product of the time $\delta$-correlated
Gaussian field $\boldsymbol{f}^{\prime}$ with other functionals in (\ref{ap:gf:v})
according to Stratonovich convention in order to preserve ordinary calculus. 
Using (\ref{ap:gf:causal}) we can write
\begin{eqnarray}
\label{}
\lefteqn{
\prec\,e^{\boldsymbol{\jmath}\star\boldsymbol{v}} \frac{\delta (\boldsymbol{\jmath}\star\boldsymbol{v})}
{\delta \bar{\boldsymbol{\jmath}}} \star (\m\partial_{\m}\kappa_{\m}\,R)\star\partial^{2}\boldsymbol{v}\,\succ
}
\nonumber\\&&
=\prec\,\frac{\delta\,e^{\boldsymbol{\jmath}\star\boldsymbol{v}} }{\delta \bar{\boldsymbol{\jmath}}}
 \star (\m\partial_{\m}\kappa_{\m}\,R)\star\partial^{2}\boldsymbol{v}\,\succ
\nonumber\\&&
=
\tr (\m\partial_{\m}\kappa_{\m}\,R)\star\partial^{2}\mathcal{Z}_{(\boldsymbol{\jmath},\bar{\boldsymbol{\jmath}})}^{(1,1)}
\ . 
\end{eqnarray}
Furthermore, a functional integration by parts yields
\begin{eqnarray}
\label{}
\lefteqn{
\prec\,e^{\boldsymbol{\jmath}\star\boldsymbol{v}} \frac{\delta (\boldsymbol{\jmath}\star\boldsymbol{v})}
{\delta \bar{\boldsymbol{\jmath}}} \star (\m\partial_{\m}\,\boldsymbol{f}^{\prime})\,\succ
}
\nonumber\\&&
=\frac{1}{2}\prec\,(\m\partial_{\m}\,\mathsf{F}^{\prime})\star\,
\frac{\delta^{2}\,e^{\boldsymbol{\jmath}\star\boldsymbol{v}} }{\delta \bar{\boldsymbol{\jmath}}
\delta \bar{\boldsymbol{\jmath}}}\succ \ ,
\end{eqnarray}
the factor $1/2$ being a consequence of Stratonovich convention.

\subsection{Ward identity}
\label{ap:gf:ward}

Let $\boldsymbol{r}_{t}:\mathbb{R}\to\mathbb{R}^{d}$ a smooth path.
The generalized Galilean transformation
\begin{subequations}
\begin{eqnarray}
\label{}
\tilde{\boldsymbol{x}}=\boldsymbol{x}+\varepsilon \,\boldsymbol{r}_{t} \ ,
\end{eqnarray}
\begin{eqnarray}
\label{}
\tilde{\boldsymbol{v}}=\boldsymbol{v}+\varepsilon \,\dot{\boldsymbol{r}}_{t} \ ,
\end{eqnarray}
\end{subequations}
leaves (\ref{intro:NS}) invariant in form when if accompanied by the
redefinition of the forcing
$\tilde{\boldsymbol{f}}=\boldsymbol{f}+\varepsilon\,\ddot{\boldsymbol{r}}_{t}$.
We must have therefore
\begin{eqnarray}
\label{}
\mathcal{Z}_{(\boldsymbol{\jmath},\bar{\boldsymbol{\jmath}})}^{(\varepsilon)}=\mathcal{Z}_{(\boldsymbol{\jmath},\bar{\boldsymbol{\jmath}})} \ .
\end{eqnarray}
When we differentiate this equality at $\varepsilon$ equal zero and
use (\ref{ap:gf:rg:var}) we obtain after standard manipulations (see e.g.
\cite{Zinn})
\begin{eqnarray}
\label{}
\lefteqn{
0=\ddot{\boldsymbol{r}}\star\left( 
\frac{\delta\,\mathcal{W}_{\left(\boldsymbol{\jmath},\bar{\boldsymbol{\jmath}}\right)}}{\delta\boldsymbol{\bar{\jmath}}}\right)
+ }
\nonumber\\&&
\hspace{-0.3cm}\boldsymbol{\jmath}\star \left(\boldsymbol{r}\cdot\boldsymbol{\partial}
\frac{\delta\,\mathcal{W}_{\left(\boldsymbol{\jmath},\bar{\boldsymbol{\jmath}}\right)}}
{\delta \boldsymbol{\jmath}}-\dot{\boldsymbol{r}}\right)
+ \boldsymbol{\bar{\jmath}}\star\,
\left(\boldsymbol{r}\cdot\boldsymbol{\partial} 
\frac{\delta\,\mathcal{W}_{\left(\boldsymbol{\jmath},\bar{\boldsymbol{\jmath}}\right)}}{\delta\boldsymbol{\bar{\jmath}}}
\right) \ .
\end{eqnarray}
An alternative way to derive the results of this appendix is based on the
Janssen--De Dominicis \cite{De76,Ja76} path integral representation of
(\ref{RGEQ:gf}). We refer to \cite{CoTo97} for a detailed presentation.

\section{Janssen--De Dominicis path integral and optimal fluctuation}
\label{ap:JDD}

The Janssen--De Dominicis \cite{De76,Ja76} representation is the formal measure
on path space obtained by requiring through an infinite dimensional product of 
Dirac $\delta$-functions that at any space-time point (\ref{intro:NS}) be satisfied.
The resulting expression is then averaged over the realizations of the stochastic 
forcing. We obtain
\begin{subequations}
\label{JDD:JDD}
\begin{eqnarray}
\label{JDD:pi}
\mathcal{Z}_{(\boldsymbol{\jmath},\bar{\boldsymbol{\jmath}})}=\int D[\boldsymbol{v}]D[\bar{\boldsymbol{v}}] e^{-\mathcal{A}} \ ,
\end{eqnarray}
\begin{eqnarray}
\label{JDD:action}
\lefteqn{
\mathcal{A}=
\frac{\bar{\boldsymbol{v}}\star\mathsf{F}\star\bar{\boldsymbol{v}}}{2}-\boldsymbol{\jmath}\star\boldsymbol{v}
}
\nonumber\\&&
-\imath\bar{\boldsymbol{v}}\star[(\partial_{t}-\kappa\partial_{\boldsymbol{x}}^{2})\boldsymbol{v}
+\mathsf{T}(\boldsymbol{v}\cdot\partial_{\boldsymbol{x}}\boldsymbol{v})-\bar{\boldsymbol{\jmath}}] \ .
\end{eqnarray}
\end{subequations}
A precise meaning to (\ref{JDD:JDD}) can be given on a space-time 
lattice using a pre-point discretization 
$dt\,(\bar{\boldsymbol{v}}\cdot\partial_{t}\boldsymbol{v})\sim \bar{\boldsymbol{v}}(t_{i})\cdot
[\boldsymbol{v}(t_{i+1})-\boldsymbol{v}(t_{i})]$, $dt\,f(\bar{\boldsymbol{v}}(t),\boldsymbol{v}(t))\sim 
dt\,f(\bar{\boldsymbol{v}}(t_{i}),\boldsymbol{v}(t_{i}))$ for all other terms in (\ref{JDD:action}). 
Notice that in the limit of vanishing stirring $\mathsf{F}\downarrow 0$, (\ref{JDD:JDD})
recovers the Fourier representation of a product of Dirac $\delta$-functions
localizing the measure over the deterministic decaying dynamics.
In this sense (\ref{JDD:JDD}) remains meaningful also as a formal
measure inclusive of compressible fluctuations.
>From (\ref{JDD:action}) a stationary phase approximation yields
the weak noise limit of the free energy $\mathcal{W}_{(\boldsymbol{\jmath},\bar{\boldsymbol{\jmath}})}$ around an optimal fluctuation
$\boldsymbol{v}^{*}$. As usual \cite{Erdelyi}, the stationary phase condition is
derived by closing a contour in the complex variables 
\begin{eqnarray}
\label{}
\bar{\boldsymbol{v}}=\bar{\boldsymbol{v}}_{\Re}+\imath\bar{\boldsymbol{v}}_{\Im} \ ,
\end{eqnarray}
which decomposes (\ref{JDD:action}) into the real and imaginary parts
\begin{subequations}
\begin{eqnarray}
\label{}
\lefteqn{\hspace{-0.7cm}
\Re\mathcal{A}_{(\boldsymbol{\jmath},\bar{\boldsymbol{\jmath}})}=
\frac{\bar{\boldsymbol{v}}_{\Re}\star\mathsf{F}\star\bar{\boldsymbol{v}}_{\Re}}{2}
-\boldsymbol{\jmath}\star\boldsymbol{v}+
}
\nonumber\\&&
\hspace{-0.6cm}
\bar{\boldsymbol{v}}_{\Im}\star\left\{(\partial_{t}
-\kappa\partial_{\boldsymbol{x}}^{2})\boldsymbol{v}+\mathsf{T}(\boldsymbol{v}\cdot\partial_{\boldsymbol{x}}\boldsymbol{v})
-\frac{1}{2}\mathsf{F}\star\bar{\boldsymbol{v}}_{\Im}-\bar{\boldsymbol{\jmath}}\right\} \ ,
\end{eqnarray}
\begin{eqnarray}
\label{}
\lefteqn{\hspace{-0.7cm}
\Im\mathcal{A}_{(\boldsymbol{\jmath},\bar{\boldsymbol{\jmath}})}
=
}
\nonumber\\&&
\hspace{-0.6cm}
-\bar{\boldsymbol{v}}_{\Re}\star\left\{(\partial_{t}
-\kappa\partial_{\boldsymbol{x}}^{2})\boldsymbol{v}
+\mathsf{T}(\boldsymbol{v}\cdot\partial_{\boldsymbol{x}}\boldsymbol{v})
-\mathsf{F}\star\bar{\boldsymbol{v}}_{\Im}
-\bar{\boldsymbol{\jmath}}\right\} \ .
\end{eqnarray}
\end{subequations}
The stationary phase condition $\Im\mathcal{A}_{(\boldsymbol{\jmath},\bar{\boldsymbol{\jmath}})}=0$ 
can then be solved for $\bar{\boldsymbol{v}}_{\Im}$ and leaves with a 
convex functional of the principal field $\boldsymbol{v}$.
Assuming that we can minimize such functional
for some assigned boundary condition, we find
within logarithmic accuracy
\begin{eqnarray}
\label{}
\lefteqn{
\mathcal{W}_{(\boldsymbol{\jmath},\bar{\boldsymbol{j}})}\sim
\boldsymbol{\jmath}\star\boldsymbol{v}^{*}
}
\nonumber\\&&
-\frac{\parallel(\partial_{t}
-\kappa\partial_{\boldsymbol{x}}^{2})\boldsymbol{v}^{*}
+\mathsf{T}(\boldsymbol{v}\cdot\partial_{\boldsymbol{x}}\boldsymbol{v})^{*}
-\bar{\boldsymbol{\jmath}}\parallel_{\mathsf{F}}^{2}}{2} \,
\end{eqnarray}
where $\parallel\boldsymbol{v}\parallel_{\mathsf{F}}^{2}$ stands for $\parallel\boldsymbol{v}\parallel_{\mathsf{F}}^{2}=\boldsymbol{v}\star\mathsf{F}^{-1}\star\boldsymbol{v}$.
The Legendre transform gives the conditions
\begin{subequations}
\begin{eqnarray}
\label{}
\boldsymbol{u}=\boldsymbol{v}^{*} \ ,
\end{eqnarray}
\begin{eqnarray}
\label{ap:JDD:inv}
\bar{\boldsymbol{u}}=\mathsf{T}\star\mathsf{F}^{-1}\star
\left\{(\partial_{t}
-\kappa\partial_{\boldsymbol{x}}^{2})\boldsymbol{v}^{*}
+\mathsf{T}(\boldsymbol{v}\cdot\partial_{\boldsymbol{x}}\boldsymbol{v})^{*}
-\bar{\boldsymbol{\jmath}}\right\} 
\end{eqnarray}
\end{subequations}
whence we finally obtain (\ref{RGEQ:tpini}). It must be stressed here
that the ``measure'' $D[\boldsymbol{v}]D[\bar{\boldsymbol{v}}]$ in (\ref{JDD:JDD}) does not exist in any rigorous mathematical 
sense. Thus, the above calculation is only formal. We give it a meaning in the following sense. 
A Gaussian measure is fully specified by its first and second moments. Since $\mathsf{F}$
is an incompressible correlation function it is consistent to consider the fields
 $\bar{\boldsymbol{v}},\bar{\boldsymbol{\jmath}}$ incompressible by definition. The field $\boldsymbol{v}^{*}$ is also
incompressible because is solution of the classical Navier--Stokes equation with vanishing 
initial condition at time $t=-\infty$ and sustained by an incompressible forcing. Finally,
the inversion operation in (\ref{ap:JDD:inv}) makes sense only away from the kernel
of the transverse correlation $\mathsf{F}$ which therefore implies that 
$\bar{\boldsymbol{u}}$ is also incompressible. 

\section{Explicit expression of the convolutions}
\label{ap:cvs}

An alternative derivation of the renormalization group equations is obtained if we observe
that  we may interpret the free energy defined by the Ansatz for the average action (\ref{eqs:tp}) 
as solution of a formal Janssen-De Dominicis \cite{De76,Ja76} path integral
\begin{eqnarray}
\label{ap:JdeDo}
\mathcal{W}(\boldsymbol{\jmath}\,,\boldsymbol{\bar{\jmath}})=\lim_{\varepsilon\searrow 0}\varepsilon\,\ln
\int D[\boldsymbol{u}]D[\boldsymbol{\bar{u}}]\,e^{
\frac{\boldsymbol{\jmath}\star\boldsymbol{u}+
\boldsymbol{\bar{\jmath}}
\star \boldsymbol{\bar{u}}-\mathcal{U}(\boldsymbol{u}\,,\boldsymbol{\bar{u}})}{\varepsilon}} \ .
\end{eqnarray}
Computing the right hand side in a perturbative expansion in powers of the interaction vertex
(\ref{approx:int1}),(\ref{approx:int2}) we obtain by standard diagrammatic techniques
\begin{eqnarray}
\label{ap:cvs:eddy}
\lefteqn{\kappa_{\m}\, p^{2}\,\gamma^{(1,1)}(p/\m)=}
\nonumber\\&&
\int \frac{ d^{d}k}{(2\,\pi)^{d}}
\frac{\left(1-\phi^{2}\right)\,
N^{(1,1)}(p,k,\phi)
\,g^{(0,2)}(k)}{2\,g^{(1,1)}(k)\,D_{1}(p,k,\phi)} \ ,
\end{eqnarray}
and
\begin{eqnarray}
\label{ap:cvs:force}
\lefteqn{\hspace{-0.4cm}
[\lambda_{(0)}\m^{2-d-2\,\varepsilon}+\lambda_{(1)}]p^{2}\gamma^{(0,2)}(p/\m)=
}
\nonumber\\&&
\hspace{-0.9cm}\int \frac{d^{d}k}{(2\,\pi)^{d}}
\frac{\left(1-\phi^{2}\right)
\,N^{(0,2)}(p,k,\phi)\, 
\,g^{(0,2)}(Q)\,g^{(0,2)}(k)}
{4\,g^{(1,1)}(k)\,
g^{(1,1)}(Q)\,D_{1}(p,k,\phi)} \ .
\end{eqnarray}
We recover equations (\ref{eqs:eqs}) by taking the logarithmic derivative
$\m\partial_{\m}$ of both sides of (\ref{ap:cvs:eddy}), (\ref{ap:cvs:force})
Note that in (\ref{ap:cvs:eddy}), (\ref{ap:cvs:force}) we denoted 
\begin{eqnarray}
\label{}
\boldsymbol{Q}:=\boldsymbol{p}-\boldsymbol{k} \ ,
\end{eqnarray}
and $\phi$ the cosine between the external $\boldsymbol{p}$ and the integration 
$\boldsymbol{k}$ wave-numbers:
\begin{eqnarray}
\label{}
\phi:=\frac{\boldsymbol{p}\cdot\boldsymbol{k}}{p\,k} \ .
\end{eqnarray}
We also defined the auxiliary integrand factors
\begin{eqnarray}
\label{}
D_{1}(p,k,\phi)=k^{2}\,g^{(1,1)}(k)+Q^{2}\,g^{(1,1)}(Q) \ ,
\end{eqnarray}
\begin{eqnarray}
\label{}
D_{2}(p,k,\phi)=2\,k^{2}\,g^{(1,1)}(k)+Q^{2}\,g^{(1,1)}(Q) \ ,
\end{eqnarray}
and the constants
\begin{eqnarray}
\label{}
C_{d}^{-1}=(d-1)\,\int_{-1}^{1}d\phi\,(1-\phi^{2})^{\frac{d-3}{2}} \ .
\end{eqnarray}
Finally, the convolutions depends upon certain integral kernels which stem from
the expansion up to one loop accuracy of the Ansatz average action (\ref{eqs:tp}).
These are 
\begin{subequations}
\begin{eqnarray}
\label{}
\lefteqn{\hspace{-0.7cm}
N^{(1,1)}(p,k,\phi):=
}
\nonumber\\&&
\hspace{-0.7cm}
\frac{(d-1)\,p^{3}\,(p-2 \,\phi\,k)+k^{2}\,p\,[(d-3)\, p+2\, \phi\,k]}
{k^{2}\,(p^{2}+k^{2}-2\,k\,p\,\phi)} \ ,
\end{eqnarray}
\begin{eqnarray}
\label{ap:cvs:n11t}
\lefteqn{\hspace{-0.7cm}
\tilde{N}^{(1,1)}(p,k,\phi):=
}
\nonumber\\&&
\hspace{-0.5cm}
\frac{p \,k\,[(d-1)\,p\,k-2\,(p^{2}+k^{2}-2\,p\,k\,\phi)\,\phi]}{k^{2}\,(p^{2}+k^{2}-2\,k\,p\,\phi)} \ ,
\end{eqnarray}
\end{subequations}
for the eddy diffusivity vertex, (\ref{ap:cvs:n11t}) will be needed below, 
and 
\begin{eqnarray}
\label{}
\lefteqn{
\hspace{-0.5cm}N^{(0,2)}(p,k,\phi):=
}
\nonumber\\&&
\hspace{-0.5cm}
\frac{p^{2}\,[(d-1)\,p^{2}-2\, d\,p\,k\,\phi+2\,k^2 \left(d+2\, \phi^{2}-2\right)]}
{k^{2}\,(p^{2}+k^{2}-2\,k\,p\,\phi)^{2}} \ ,
\end{eqnarray}
for the force vertex. Finally  in (\ref{eqs:eqs}) there appear terms of the form
\begin{eqnarray}
\label{ap:cvs:conv}
\lefteqn{\hspace{-1.0cm}
G^{(i,j)}_{l}(p):=
}
\nonumber\\&&\hspace{-0.7cm}
\frac{C_{d}}{2\,p^{2}}\int_{0}^{\infty} \frac{dk}{k}\,k^{d}
\int_{-1}^{1}d\phi\,(1-\phi^{2})^{\frac{d-1}{2}}\,V^{(i,j)}_{l}(p,k,\phi) \ ,
\end{eqnarray}
with $l$ taking values $\left\{F,\kappa,o\right\}$ and $V^{(i,j)}_{l}(p,k,\phi)$ the non-linear convolutions specified below.

\subsection{Equation for the eddy diffusivity vertex}
\label{ap:cvs:e}

The following three non-linear convolutions enter (\ref{eqs:eddy}): 
\begin{eqnarray}
\hspace{-1.0cm}V^{(1,1)}_{F}(p,k,\phi):=
\frac{N^{(1,1)}(p,k,\phi)\,\lambda_{(1)}\chi_{(1)}(k)}
{g^{(1,1)}(k)\,D_{1}(p,k,\phi)
} \ ,
\end{eqnarray}
with coefficient $\eta_{F}$,
\begin{eqnarray}
\lefteqn{\hspace{-0.6cm}
V^{(1,1)}_{\kappa}(p,k,\phi):=\frac{\check{R}(k)}
{[D_{1}(p,k,\phi)]^{2}
}\times
}
\nonumber\\&&
\left\{
\frac{
D_{2}(p,k,\phi)
\,N^{(1,1)}(p,k,\phi)\,g^{(0,2)}(k)}
{[g^{(1,1)}(k)]^{2}}
\right.
\nonumber\\&&
\left.
+\frac{k^{4}\,\tilde{N}^{(1,1)}(p,k,\phi)\,g^{(0,2)}(Q)}{Q^{2}g^{(1,1)}(Q)}
\right\} \ ,
\end{eqnarray}
with coefficient $\eta_{\kappa}$, and
\begin{eqnarray}\label{conv-V11}
\lefteqn{
V^{(1,1)}_{o}(p,k,\phi)=
}
\nonumber\\&&
\frac{N^{(1,1)}(p,k,\phi)\,\sum_{i=0}^{1}
\lambda_{(i)}\,(\boldsymbol{k}\cdot\partial_{\boldsymbol{k}}
-d_{F_{(i)}})\chi_{(i)}(k)}{g^{(1,1)}(k)\,D_{1}(p,k,\phi)}
\nonumber\\&&
-\frac{(\boldsymbol{k}\cdot\partial_{\boldsymbol{k}}\check{R})(k)}
{[D_{1}(p,k,\phi)]^{2}}
\left[\frac{D_{2}(p,k,\phi)\,N^{(1,1)}(p,k,\phi)\,g^{(0,2)}(k)}{g^{(1,1)}(k)}
\right.
\nonumber\\&&
\left.+\frac{k^{4}\,\tilde{N}^{(1,1)}(p,k,\phi)\,g^{(0,2)}(Q)}{Q^{2}\,g^{(1,1)}(Q)}
\right] \ ,
\end{eqnarray}
with coefficient equal to the unity.

\subsection{Equation for the force vertex}
\label{ap:cvs:f}

The following three non-linear convolutions enter (\ref{eqs:force}):
\begin{eqnarray}
\hspace{-0.7cm}
V_{F}^{(0,2)}(p,k,\phi):=
\frac{N^{(0,2)}(p,k,\phi)\,g^{(0,2)}(Q)\,\chi_{(1)}(k)}
{g^{(1,1)}(Q)\,g^{(1,1)}(k)\,D_{1}(p,k,\phi)} \ ,
\end{eqnarray}
with coefficient $\eta_{F}$,
\begin{eqnarray}
\lefteqn{
\hspace{-0.7cm}
V_{\kappa}^{(0,2)}(p,k,\phi):=N^{(0,2)}(p,k,\phi)
\times
}
\nonumber\\&&
\frac{g^{(0,2)}(Q)\,g^{(0,2)}(k)\,\check{R}(k)
D_{2}(p,k,\phi)
}{g^{(1,1)}(Q)\,[g^{(1,1)}(k)]^{2}
[D_{1}(p,k,\phi)]^{2}
} \ ,
\end{eqnarray}
with coefficient $\eta_{\kappa}$, and
\begin{eqnarray} \label{conv-V02}
\lefteqn{
\hspace{-1.0cm}
V_{o}^{(0,2)}(p,k,\phi):=
\frac{N^{(0,2)}(p,k,\phi)\,g^{(0,2)}(Q)}
{g^{(1,1)}(Q)\,g^{(1,1)}(k)\,D_{1}(p,k,\phi)
}\times
}
\nonumber\\&&
\left\{\sum_{i=0}^{1}\lambda_{(i)}\left(\boldsymbol{k}\cdot\partial_{\boldsymbol{k}}-d_{F_{(i)}}\right)
\chi_{(i)}(k,\mu)\right.
\nonumber\\&&
\left.
-\frac{(\boldsymbol{k}\cdot\partial_{\boldsymbol{k}}\check{R})(k)\,g^{(0,2)}(k)}{g^{(1,1)}(k)}
\,\frac{D_{2}(p,k,\phi)
}{D_{1}(p,k,\phi)
}\right\} \ ,
\end{eqnarray}
with coefficient equal to the unity.

\bibliographystyle{h-physrev}

\end{document}